\documentclass[twocolumn,showpacs,amsmath,amssymb,aps,pra,floatfix,10pt]{revtex4-2}
\usepackage[latin1]{inputenc}
\usepackage[english]{babel}
\usepackage[T1]{fontenc}
\usepackage{mathrsfs}
\usepackage{hyperref}
\usepackage{multirow}
\usepackage{bm,epsfig}
\usepackage{graphicx}
\usepackage{subeqnarray}
\usepackage{bbold}
\usepackage{upgreek}
\hypersetup{colorlinks=true, urlcolor=blue, citecolor=blue, filecolor=blue, linkcolor=blue}
\usepackage{dcolumn}
\usepackage{amsmath}
\usepackage{color}

\renewcommand{\Re}{\operatorname{Re}}
\renewcommand{\Im}{\operatorname{Im}}

\begin{document}
\title{Polarisation control of quasi-monochromatic XUV \\produced via resonant high harmonic generation
}
\author{M.~A.~Khokhlova$^{1,\ast}$, M.~Yu.~Emelin$^{2}$, M.~Yu.~Ryabikin$^{2}$, and V.~V.~Strelkov$^{3,4}$
}
\affiliation{
\mbox{$^1$Max Born Institute for Nonlinear Optics and Short Pulse Spectroscopy, Max-Born-Stra{\ss}e 2A, Berlin 12489, Germany} \\
\mbox{$^2$Institute of Applied Physics of the Russian Academy of Sciences, 46 Ulyanov street, Nizhny Novgorod 603950, Russia} \\
\mbox{$^3$Prokhorov General Physics Institute of the Russian Academy of Sciences, 38 Vavilova street, Moscow 119991, Russia} \\ 
\mbox{$^4$Moscow Institute of Physics and Technology (State University), Dolgoprudny, Moscow Region 141700, Russia} \\
$^{\ast}$m.khokhlova@imperial.ac.uk
}

\date{\today}

\begin{abstract}\noindent 
We present a numerical study of the resonant high harmonic generation by tin ions in an elliptically-polarised laser field along with a simple analytical model revealing the mechanism and main features of this process. We show that the yield of the resonant harmonics behaves anomalously with the fundamental field ellipticity, namely the drop of the resonant harmonic intensity with the fundamental ellipticity is much slower than for high harmonics generated through the nonresonant mechanism. Moreover, we study the polarisation properties of high harmonics generated in elliptically-polarised field and show that the ellipticity of harmonics near the resonance is significantly higher than for ones far off the resonance. This introduces a prospective way to create a source of the quasi-monochromatic coherent XUV with controllable ellipticity potentially up to circular. 
\end{abstract}

\maketitle
\noindent

\section{Introduction}
Extreme ultraviolet (XUV) sources of ultrafast emission have proven themselves to be an immensely important and powerful tools for tracking and control of electron dynamics in atoms, molecules, and condensed matter~\cite{atto_rev,femto_rev}. One of the handles of these tools is the polarisation of the emitted electromagnetic field. In particular, circularly-polarised (CP) XUV pulses recently have become extremely valuable due to their broad involvement in a growing number of experimental techniques for the study of structural, electronic, and magnetic properties of matter, such as chiral molecules~\cite{PhysRevLett.86.1187,Beaulieu2016Dec} and magnetic materials~\cite{Boeglin2010May,Kfir2017Dec,PhysRevB.92.220405,Siegrist2019Jun}, as well as for deeper investigation of strong-field processes in atoms (such as nonsequential photoionisation)~\cite{PhysRevA.99.023413} or other curious applications~\cite{Prost2018Nov}. 

On the one hand, although synchrotrons~\cite{Boeglin2010May} and free-electron lasers~\cite{PhysRevX.4.041040} are typically used as a source of CP XUV pulses and perform some progress, they still suffer from the low degree of coherence. An alternative group of sources, which does not only allow to switch from large-scale facilities towards table-top setups, but also enables the significant improvement of the coherence properties of the emission, is based on the high-order harmonic generation (HHG) process~\cite{atto_rev}. 

The naive approach to create CP harmonics~--- via HHG by atoms in an elliptically-polarised (EP) field~--- fails immediately~\cite{Budil_PRA_48}. Although the ellipticity of high harmonics grows with the driver's (fundamental) ellipticity~\cite{Antoine_Lew}, there is a limiting factor: the harmonic yield drops rapidly (exponentially) with the fundamental ellipticity~\cite{Budil_PRA_48,Dietrich1994,PhysRevA.86.011401}. Thus, HHG in an atomic gaseous medium driven by an EP laser field does not appear as a reasonable candidate for EP XUV sources. 

There have been extensive efforts towards the production of EP and CP XUV via HHG recently, resulting in the suggestion of numerous nontrivial experimental schemes. One class of such schemes involves bicircular bichromatic fields~{\cite{Eichmann1995,PhysRevA.52.2262,Emilio_Misha,Fleischer:14,Kfir2014Dec,Fan2015Nov} or other geometries of two-colour fields~\cite{Lambert_bicircular,Murnane_PRA,Bandrauk_bicircular}. A second class manipulates the polarisation of the generated XUV using reflection-~\cite{Vodungbo2011Feb,PhysRevLett.115.083901,PhysRevB.92.220405,Siegrist2019Jun} and transmission-~\cite{Schmidt2015Dec} based polarisers. Other recent demonstrations include various complex combinations of multiple beams or multiple targets which produce EP emission through delicate interference between different components~\cite{Azoury2019Feb,Ellis2018,Huang2018}. One more way to produce EP XUV pulses is HHG using aligned molecules in gas phase~\cite{PhysRevLett.102.073902,PhysRevLett.104.213601,Skantzakis2016Dec}. These schemes rely on rather sophisticated experimental setups, which make them harder to implement. Also the common drawback of these schemes, as of based on HHG, is a low intensity of emitted XUV.

In this paper we suggest a new way to generate bright XUV with tunable polarisation via HHG more efficiently, avoiding overcomplicated experimental schemes. Our approach is based on the process of resonant HHG~\cite{Ganeev2006Jun,PhysRevLett.102.243901,Strelkov_PRL_104,Shiner2011,Strelkov_Fano}: if the harmonic energy is close to one of the transitions between the ground state and an autoionising state (AIS) of the generating particle, with a high oscillator strength, then the intensity of this `resonant' harmonic becomes significantly boosted, by up to a couple of orders of magnitude. Since the AIS is much less localised, in contrast to the ground state, it becomes possible for the electron in an EP laser field to be captured into the AIS for higher fundamental ellipticities than in the nonresonant case, with the subsequent recombination to the ground state accompanied by the enhanced XUV emission. Moreover, resonant harmonics can carry higher ellipticities, which make them a prospective candidate for a source of quasi-monochromatic EP or even CP XUV radiation with relatively high intensity. Our approach allows efficient generation of XUV with high ellipticity in a potentially wide spectral range, in contrast to similar approaches to generate quasi-CP XUV, using excited (or Rydberg) states~\cite{Ferre2014_rhhg,Beaulieu2016,Camp2018} or shape resonances~\cite{Ferre2015Jan} instead of AIS, which are limited to either near-threshold harmonics or appears to be a potentially much weaker effect, respectively.

To substantiate our proposed scheme, we study numerically the resonant HHG in an EP laser field by solving the three-dimensional time-dependent Schr\"{o}dinger equation (3D TDSE) for the singly ionised tin atom (SnII) in a laser field of different wavelengths and varying ellipticity. Our results show that the efficiency of the resonant harmonics exhibits much slower decrease with the fundamental ellipticity than the efficiency of nonresonant harmonics~\cite{Budil_PRA_48,Dietrich1994,Antoine_PRA_53}. Moreover, the resonant harmonic yield sometimes shows anomalous behaviour, which is expressed in its constancy or even growth for some laser parameters. We also study the behaviour of the harmonic ellipticity, which shows the strong impact of the resonance on it. Not only the ellipticity of the resonant harmonic itself appears to be rather high, even close to unity, but the ellipticity of nonresonant harmonics is affected by the resonance, including the change of the ellipticity sign as well. These numerical results lay in line with our analytical toy model.

\section{Numerical methods}\label{nummethods}
We study numerically the resonant HHG in an EP laser field by solving fully 3D TDSE in a single-active electron (SAE) approximation. 

The TDSE was integrated numerically using the fast Fourier transform-based split-operator technique~\cite{Fleck1976}. The calculations were performed using a multithreaded numerical code we have created on the basis of libraries implementing the POSIX Threads standard.

The model potential reproducing the interaction of the active electron with the nucleus and with the rest of the electrons is chosen in the form suggested in~\cite{Strelkov_PRL_104}. It can be written as a combination of a soft-core Coulomb potential and a barrier, thus allowing for a quasi-stable state with positive energy, which models the AIS:
\begin{equation}
V(r)=-\frac{Q+1}{\sqrt{a_0^2+r^2}}+a_1 \exp{\left[-\left(\frac{r-a_2}{a_3} \right)^2 \right]} \, .
\label{mod_pot}
\end{equation}
Here $Q=1$ (atomic units are used if not stated otherwise) is the charge state of the generating particle, and $a_0=0.66$, $a_1=1.2$, $a_2=3.56$, $a_3=2.5$ are constants chosen to replicate the properties of the generating particle (SnII): the ground state energy (or $-E_\mathrm{ground}=I_\mathrm{p}=14.63$~eV), the resonant energy ($\Delta E=E_\mathrm{AIS}-E_\mathrm{ground}=26.2$~eV), the AIS width ($\Gamma_\mathrm{AIS}=0.16$~eV) and the oscillator strength ($gf=1.43$) of the transition between the ground state and the AIS (see~\cite{ganeev_pra85} for details). 

The laser pulse $E(t)=E_0 f(t) \sin{(\omega_0 t)}$ with frequency $\omega_0$ and electric field amplitude $E_0$ used for our calculations has a trapezoidal envelope 
\begin{small}
\begin{equation*}
f(t)=
\begin{cases} 
\sin^2{\left( \frac{\pi}{2} \frac{t}{\tau_f} \right)} \, , & t \leq \tau_f \\
1 \, , & \tau_f < t < \tau_f + \tau_c \\
\sin^2{\left( \frac{\pi}{2} \frac{t-(2\tau_f+\tau_c)}{\tau_f} \right)} \, , & \tau_f + \tau_c \leq t \leq 2\tau_f + \tau_c \\
0 \, , & 2\tau_f + \tau_c < t < 2\tau_f + \tau_c + \tau_z
\end{cases}
\end{equation*}
\end{small}%
with 4 cycles of constant intensity, $\tau_c$, and 2 cycles of turning-on and turning-off each, $\tau_f$. We extend the calculation time by the AIS lifetime (corresponding to $\Gamma_\mathrm{AIS} \sim \tau_z$) to account for the longer emission of the resonant harmonic.

From the TDSE solution, we obtain time-dependent dipole moment components $d_x(t)$ and $d_y(t)$, directed along the polarisation ellipse axes of the laser field. We then calculate the Stokes parameters $S_0$, $S_1$, $S_2$ and $S_3$,
\begin{equation}
\begin{split}
S_0=&|d_x(\omega)|^2+|d_y(\omega)|^2 \, , \\
S_1=&|d_x(\omega)|^2-|d_y(\omega)|^2 \, , \\
S_2=&2\Re\left[d_x(\omega)d^*_y(\omega)\right] \, , \\
S_3=&-2\Im\left[d_x(\omega)d^*_y(\omega)\right] \, ,
\end{split}
\label{stocks}
\end{equation}
using the spectral representation for dipole components, $d_x(\omega)$ and $d_y(\omega)$. Thus, the harmonic spectrum in this case is $S_0$, and the harmonic polarisation properties are obtained from the Stokes parameters as
\begin{equation}
\psi=\frac{1}{2}\arctan{\left(\frac{S_2}{S_1} \right)} \, , 
\label{num_an}
\end{equation}
for the rotation angle, and 
\begin{equation}
\epsilon=-\tan{\left[\frac{1}{2} \arctan{\left( \frac{S_3}{\sqrt{(S_1^2+S_2^2)}} \right)} \right]} \, ,
\label{num_ell}
\end{equation}
for the ellipticity.

\section{Analytical toy model}\label{toymodel}
The rapid decrease of the harmonic intensity with the fundamental ellipticity in the absence of resonances can be understood in the framework of the three-step model~\cite{3-step_C,3-step_S,3-step_K}: in an EP field, the electron wavepacket returning to the parent ion is transversely shifted relative to the parent ion, and at some threshold ellipticity~\cite{Budil_PRA_48,Strelkov_PRA_EPL} it starts missing the parent ion completely. 

\begin{figure}[ht]\centering 
\includegraphics[width=1.0\columnwidth]{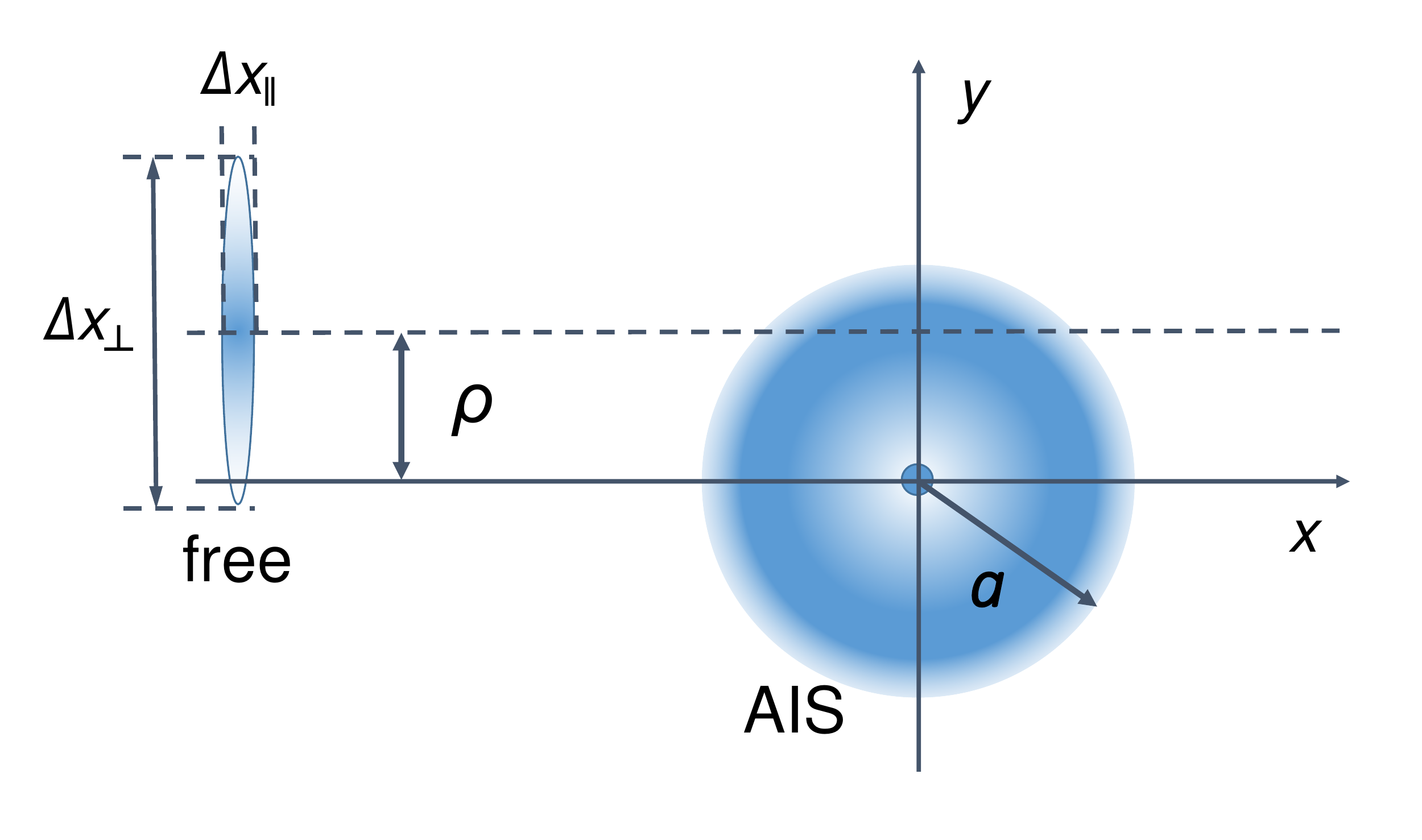}
\caption{Illustration of the model. The diffused free electronic wavepacket $\Psi_\mathrm{free}$ ($\Delta x_\mathrm{\parallel}$ and $\Delta x_\mathrm{\perp}$ are its uncertainties) moves back towards the parent ion, which has an AIS of effective size $a$, with the impact parameter $\rho$.}
\label{fig:scheme}
\end{figure}
In the case of resonant HHG, the mechanism behind the anomalous dependence of the resonant harmonic on the fundamental ellipticity can be explained using the four-step model~\cite{Strelkov_PRL_104}. The first two steps (electron ionisation and propagation of the electron wavepacket in the driving field) follow the three-step model, but instead of immediate radiative recombination into the ground state, the returning electron gets captured into an AIS of the parent ion (new third step), from which it then relaxes into the ground state emitting the XUV photon (fourth step). Since the AIS is much less tightly localised than the ground state, an electron wavepacket, which would miss the ground state in nonresonant HHG, continues to hit the parent ion AIS at much higher ellipticities of the driver (see Fig.~\ref{fig:scheme}). This leads to a slower decrease or even to an anomalous (constant or even locally increasing) behaviour of the resonant harmonic efficiency (yield) with fundamental ellipticity in comparison to nonresonant harmonics. 

Further we consider the influence of the AIS size on the HHG in an EP laser field. Formally, the contribution of the resonance to the yield $I_q$ of the resonant $q$-th order harmonic can be understood as a product of the form:
$I_q \approx w_i S R \, ,$
where $w_i$ is the ionisation probability (which can be found within the ADK~\cite{ADK} or PPT~\cite{PPT} formulae, for instance), $S$ is a factor responsible for the trapping of the free electron into the AIS (depending, in particular, on features of the free-electronic motion between the detachment and the return, see below), and $R$ is the recombination factor describing the transition from the AIS to the ground state. 

The factor $S=|\langle \Psi_\mathrm{AIS}|\Psi_\mathrm{free}\rangle|^2$, which carries the difference between the resonant and nonresonant cases, is essentially the squared absolute value of the overlap integral of the AIS wavefunction $\Psi_\mathrm{AIS}$ and the free electron wavepacket $\Psi_\mathrm{free}$ returning to the parent ion. Here $\Psi_\mathrm{AIS}$ is the spatial part of the AIS wavefunction, for simplicity, modelled by the hydrogenic set of $p$-orbitals $\Psi_\mathrm{AIS}=\sum\Psi_{21m}$, where
\begin{equation}
\begin{split}
\Psi_{210}=&\frac{1}{4 \sqrt{2 \pi}}\left( \frac{Z}{a} \right)^{5/2} e^{-\frac{Z r}{2a}} r \cos{\theta} \, ,\\
\Psi_{21\pm 1}=&\frac{1}{8 \sqrt{\pi}}\left( \frac{Z}{a} \right)^{5/2} e^{-\frac{Z r}{2a}} r \sin{\theta} e^{\pm \varphi} \, 
\end{split}
\label{wfs}
\end{equation} 
with an effective radius $a$ (see Fig.~\ref{fig:scheme}) and effective charge $Z$~\cite{Clementi1967Aug}. 
The symmetry and size of the chosen states here are the same as in our numerical SAE TDSE calculations.

The free electron wavefunction $\Psi_\mathrm{free}$, before capturing, is described as a Gaussian wavepacket (see Fig.~\ref{fig:scheme})
\begin{equation}
\begin{split}
\Psi_\mathrm{free}=&\left(\frac{1}{2 \pi \Delta x_\mathrm{\parallel}^2}\right)^{1/2} e^{-\frac{(r \sin{\theta} \sin{\varphi})^2}{2 \Delta x_\mathrm{\parallel}^2}} \frac{1}{2 \pi \Delta x_\mathrm{\perp}^2} e^{-\frac{(r \cos{\theta})^2}{2 \Delta x_\mathrm{\perp}^2}} \\
& \times e^{-\frac{(r \sin{\theta} \cos{\varphi}- \rho)^2}{2 \Delta x_\mathrm{\perp}^2}} e^{i \sqrt{2(q \omega_{0} - I_\mathrm{p})}r \sin{\theta} \sin{\varphi}} \, ,
\end{split}
\label{free}
\end{equation}
where $\Delta x_\mathrm{\parallel}$ and $\Delta x_\mathrm{\perp}$ are uncertainties of the electron wavepacket in the longitudinal and perpendicular directions (see~\cite{PPT,Strelkov_PRA_EPL} for details), correspondingly, with respect to the direction of the laser field (see Fig.~\ref{fig:scheme}), $\rho$ is the impact parameter (or the shift of the centre of the electron wavepacket with respect to the parent ion) calculated from the classical equation of motion, the wavenumber is $k=\sqrt{2(q \omega_{0} - I_\mathrm{p})}$, and $I_\mathrm{p}$ is the ionisation potential of the generating particle. 

Calculating the properties of the returning electronic wavepacket for different laser ellipticities and harmonic orders, we find the factor $S$ and thus the harmonic intensity.

\section{Results}

\subsection{Harmonic yield}
We consider the HHG by tin ions (SnII) in an EP laser field with varying ellipticity by solving TDSE numerically (see Sec.~\ref{nummethods} for details). Three different fundamental wavelengths (800~nm, 520~nm, and 1300~nm) are used, the fundamental ellipticity varies between 0 and 0.5, and the fundamental intensity in all our calculations is kept constant, $2 \cdot 10^{14}$ W/cm$^2$.

We start with the most typical fundamental Ti:Sapphire wavelength, 800~nm, where the resonant harmonic is H17. Fig.~\ref{fig:spec} demonstrates the evolution of the HHG spectrum with increasing laser ellipticity. One can see that there are two types of behaviour: the intensity of nonresonant harmonics rapidly drops down with increasing fundamental ellipticity, while the intensity of the resonant one does not decrease in this way. 

\begin{figure}[t]\centering 
\includegraphics[width=1.0\columnwidth]{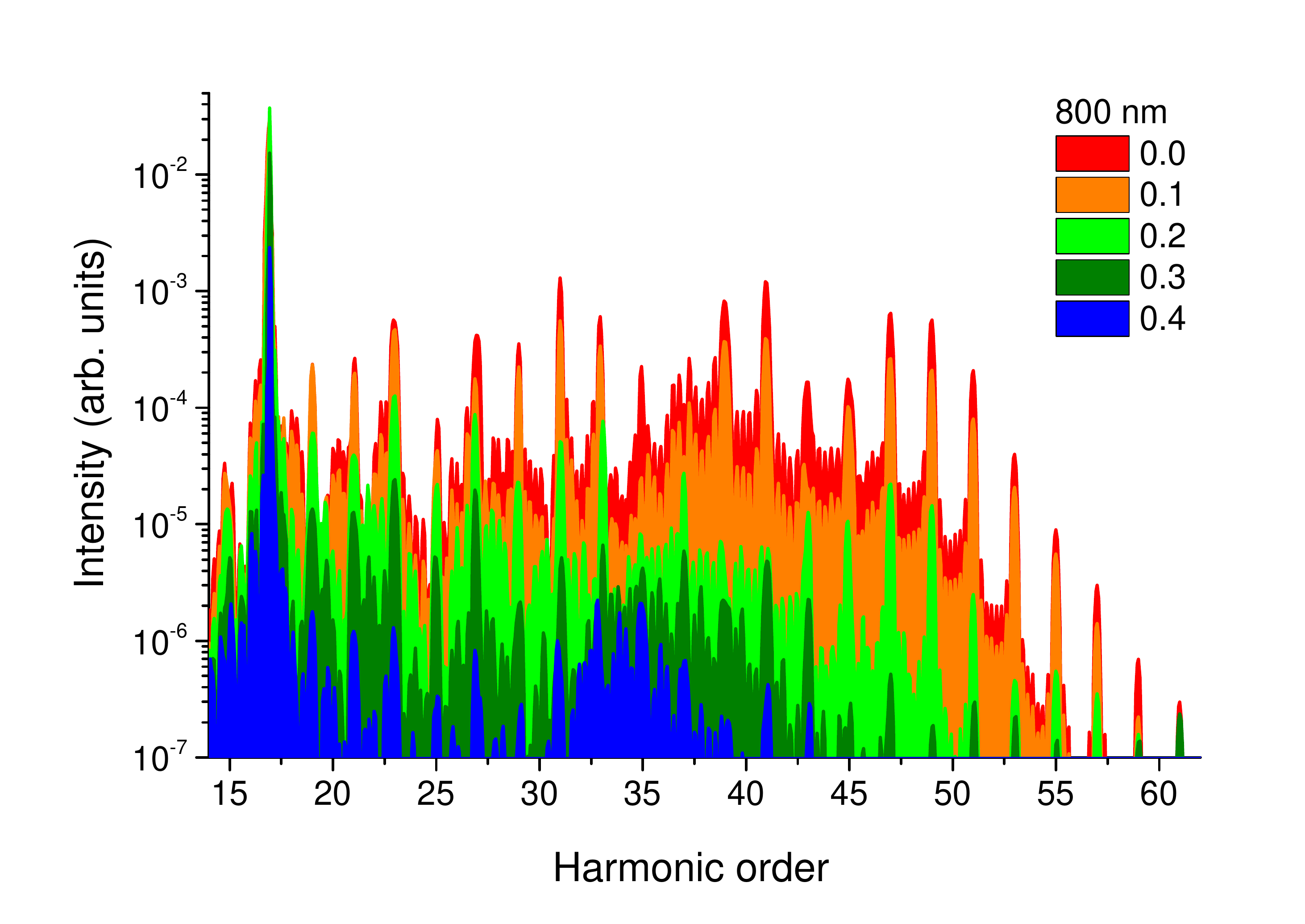}
\caption{Harmonic spectra calculated within the SAE TDSE for the fundamental wavelength 800~nm, and fundamental ellipticities from 0 to 0.4.}
\label{fig:spec}
\end{figure}

More explicitly, this difference in the behaviour of the yield for the resonant (H17) and for one of the nonresonant harmonics from the middle of the plateau (we choose H29), as a function of the fundamental ellipticity, is presented in Fig.~\ref{fig:yield}(a). Here it is shown that the yield of the nonresonant H29 harmonic decreases quickly with the fundamental ellipticity, so that at fundamental ellipticity 0.2 the H29 yield loses one order of magnitude, and then two orders for the fundamental ellipticity 0.3.

In contrast, the yield of the resonant H17 harmonic not only starts from significantly higher value for the linearly-polarised field, but remains on an almost constant level up to the fundamental ellipticity 0.3, increasing slightly at low fundamental ellipticities and reaching a local maximum at 0.15. Note that a slightly anomalous behaviour of nonresonant harmonics explained by the quantum interference effects was shown in~\cite{Antoine_PRA_53}, which is smoothened out by propagation effects. The exponential decrease of the H17 yield with the same rate as for H29 starts at the fundamental ellipticity 0.35. We note that the black diamond in Fig.~\ref{fig:yield}(a) shows the numerical result for three times more spatially dense grid, which allows to resolve the higher angular momenta appearing for high fundamental ellipticities. The convergence of the results for the lower elliplicities has been reached.

\begin{figure}[t!]\centering 
\includegraphics[width=1.0\columnwidth]{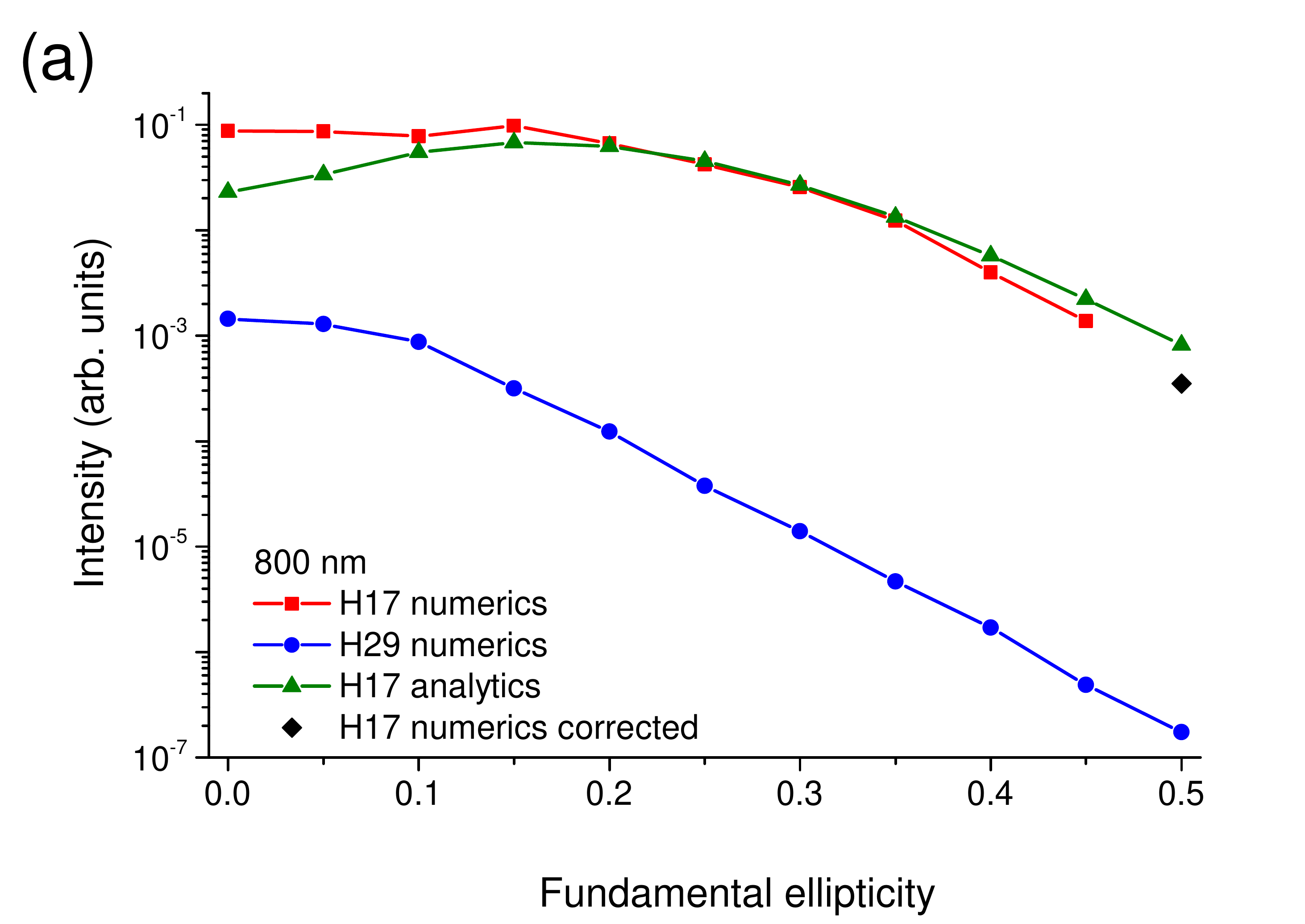}
\includegraphics[width=1.0\columnwidth]{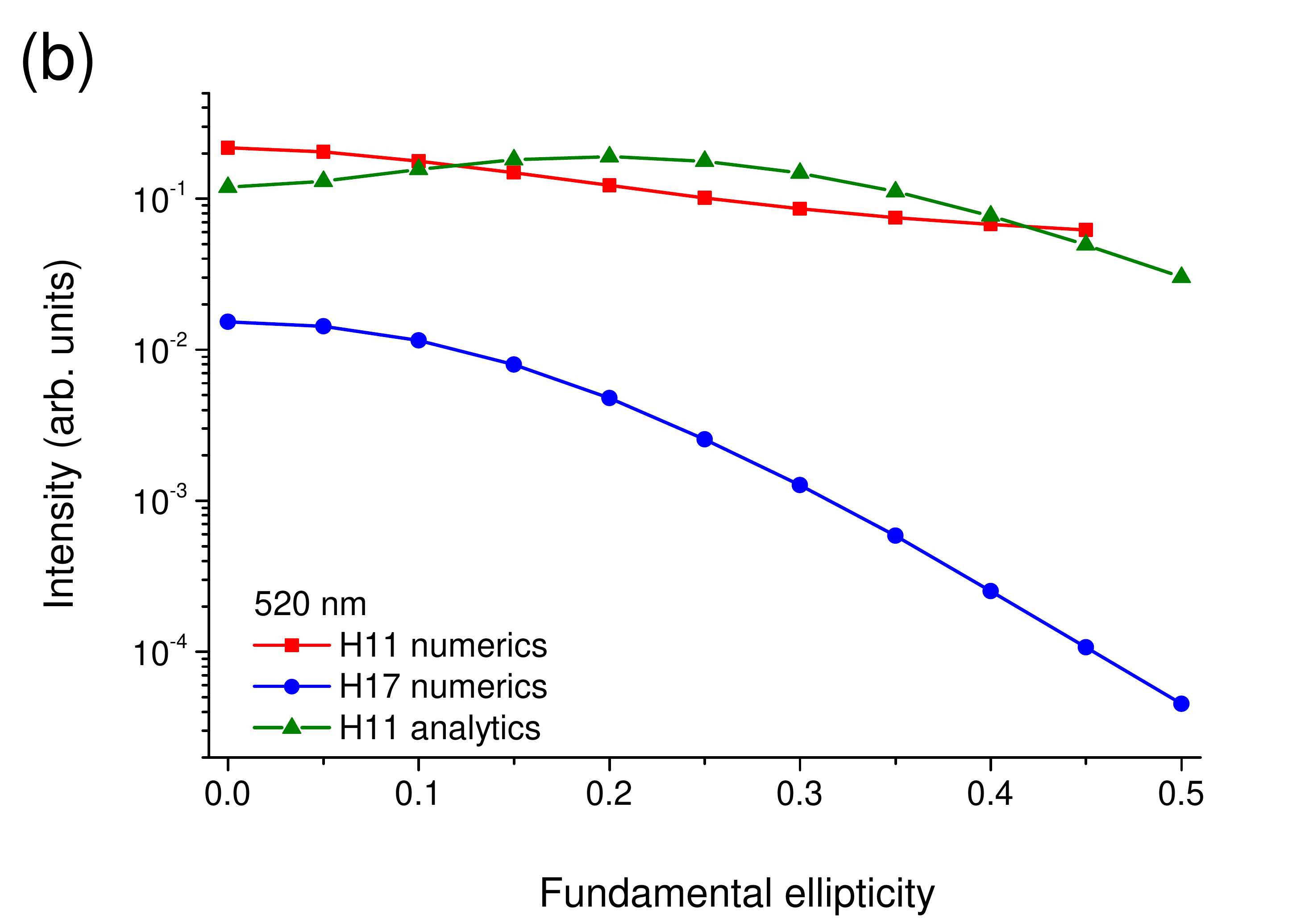}
\includegraphics[width=1.0\columnwidth]{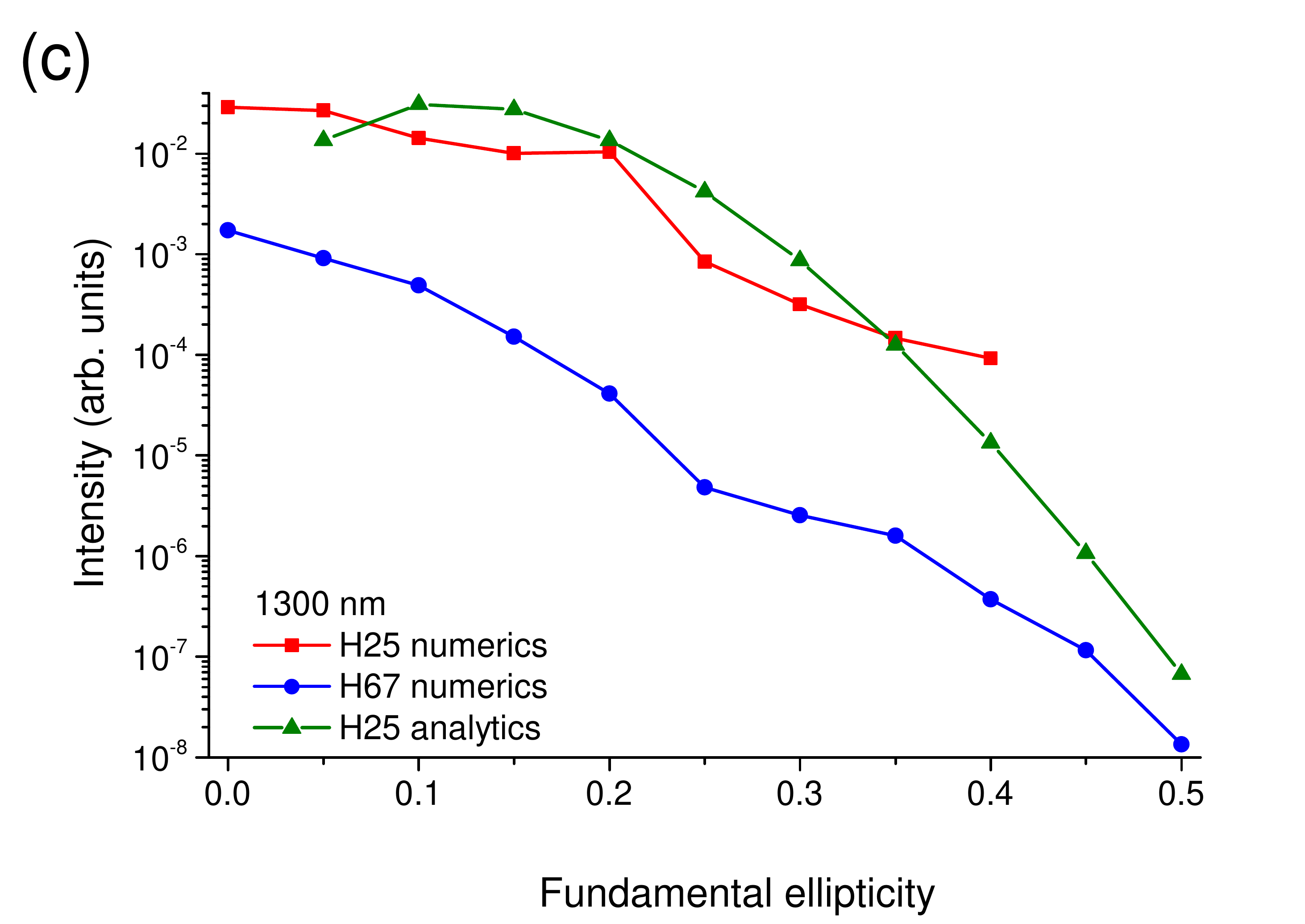}
\caption{Harmonic intensity as a function of the fundamental ellipticity for fundamental wavelengths (a) 800 nm, (b) 520 nm and (c) 1300 nm. Pairs of resonant (red squares) and one of nonresonant (blue circles) harmonic intensities calculated within the SAE TDSE are presented for each wavelength. Green triangles show analytical resonant contribution to harmonic intensity.}
\label{fig:yield}
\end{figure}

We also present the results for smaller (520~nm) and larger (1300~nm) fundamental field wavelengths in Fig.~\ref{fig:yield}, graphs (b) and (c), where H11 and H25 are the resonant harmonics, correspondingly. The behaviour of the yield in these cases is similar to the case of 800~nm, with either a slightly slower (520~nm) or slightly faster (1300~nm) decrease with increasing fundamental ellipticity; this can be explained within the recollision picture for a less or more diffused electron wavepacket, respectively. The resonant yield of the H11 in Fig.~\ref{fig:yield}(b) decreases very slowly, and for H25 in Fig.~\ref{fig:yield}(c) it even reverses its decrease briefly to reach a local maximum at ellipticity 0.2. Results for relatively high fundamental ellipticities [$>0.45/0.4$ for panels (b)/(c)] are obtained within a less dense spatial grid, and therefore are not precise and are not displayed.

We compare our numerical results with the toy model (see Sec.~\ref{toymodel}), which is shown with green triangles in Figs.~\ref{fig:yield}(a-c). The analytical results only capture the resonant (four-step) mechanism of HHG, while the full resonant harmonic yield also includes the nonresonant (three-step model) contribution. This leads to the difference between results obtained with analytical toy model and numerical simulations. However, the appearance and the position of local maximum can still be explained by the geometry of the AIS, which was chosen as the simplest geometry ($p$-orbitals) allowing the AIS~-- ground state transition.

Also the analytical model reproduces much slower decrease of the resonant harmonic intensity with the laser ellipticity for 520~nm and 800~nm fundamental and comparable decreases of resonant and non-resonant harmonics for 1300~nm fundamental wavelength.

\subsection{Harmonic polarisation properties}
Here we study the polarisation properties~--- the rotation angle and the harmonic ellipticity~--- of high harmonics, generated by tin ions in the EP field, as discussed in the previous section.

Figure~\ref{fig:rot_an} shows the rotation angle \eqref{num_an} of the polarisation ellipse of resonant and nonresonant harmonics, generated by the field with 800~nm wavelength, as a function of the harmonic order for fundamental ellipticities 0.05 to 0.2.
Here one can see that the rotation angle of harmonics around the resonance (H17) behaves irregularly with the harmonic order, and the rotation angle value becomes relatively large, while for harmonics far form the resonance, this behaviour is demonstrated to be smooth~\cite{Strelkov_PRA_EPL} and the rotation angle typically decreases for higher orders.

\begin{figure}[t]\centering 
\includegraphics[width=1.0\columnwidth]{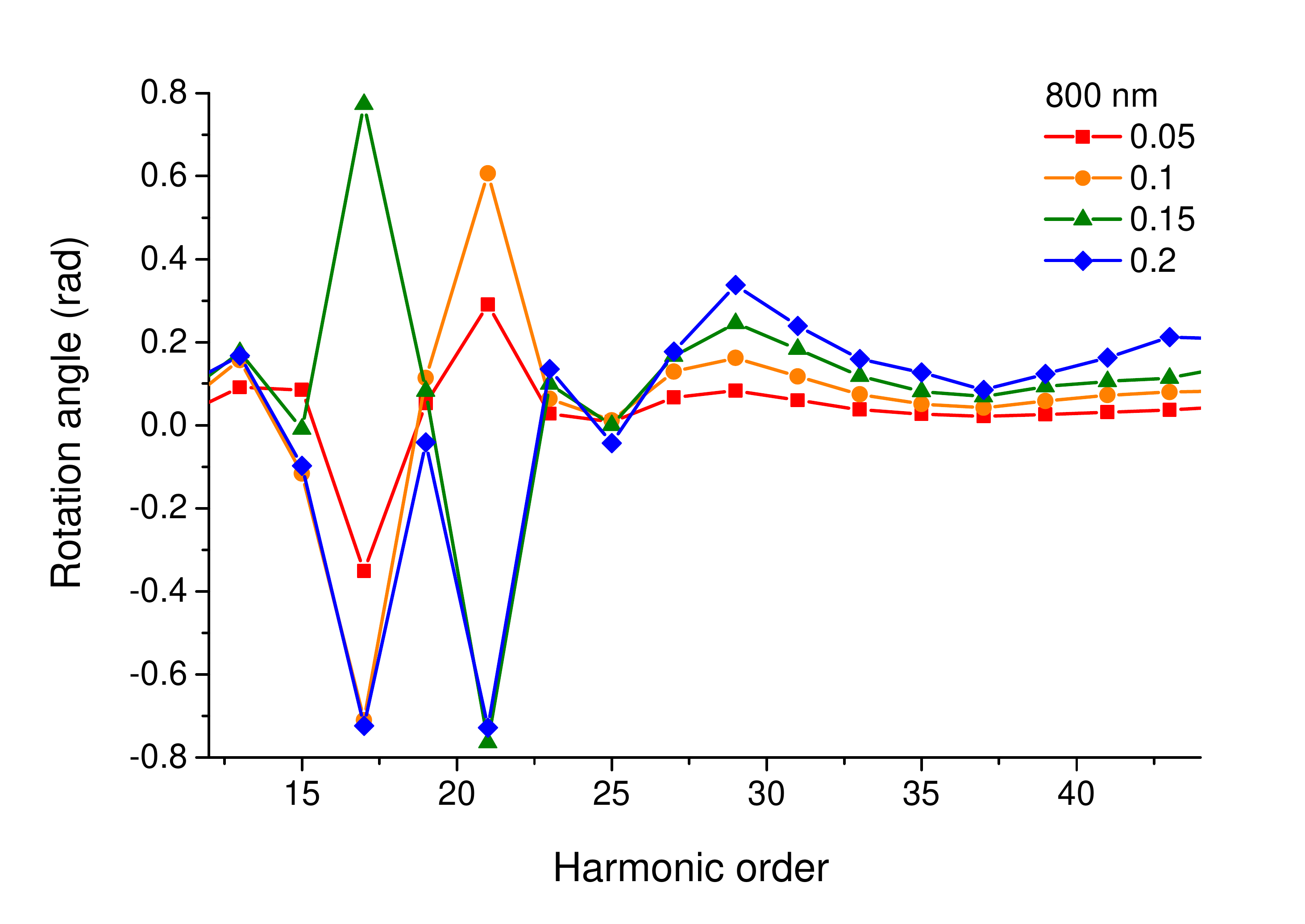}
\caption{The harmonic rotation angle as a function of the harmonic order for the fundamental wavelength 800~nm and fundamental ellipticities from 0.05 to 0.2 calculated within the SAE TDSE.}
\label{fig:rot_an}
\end{figure}

\begin{figure}[t]\centering 
\includegraphics[width=1.0\columnwidth]{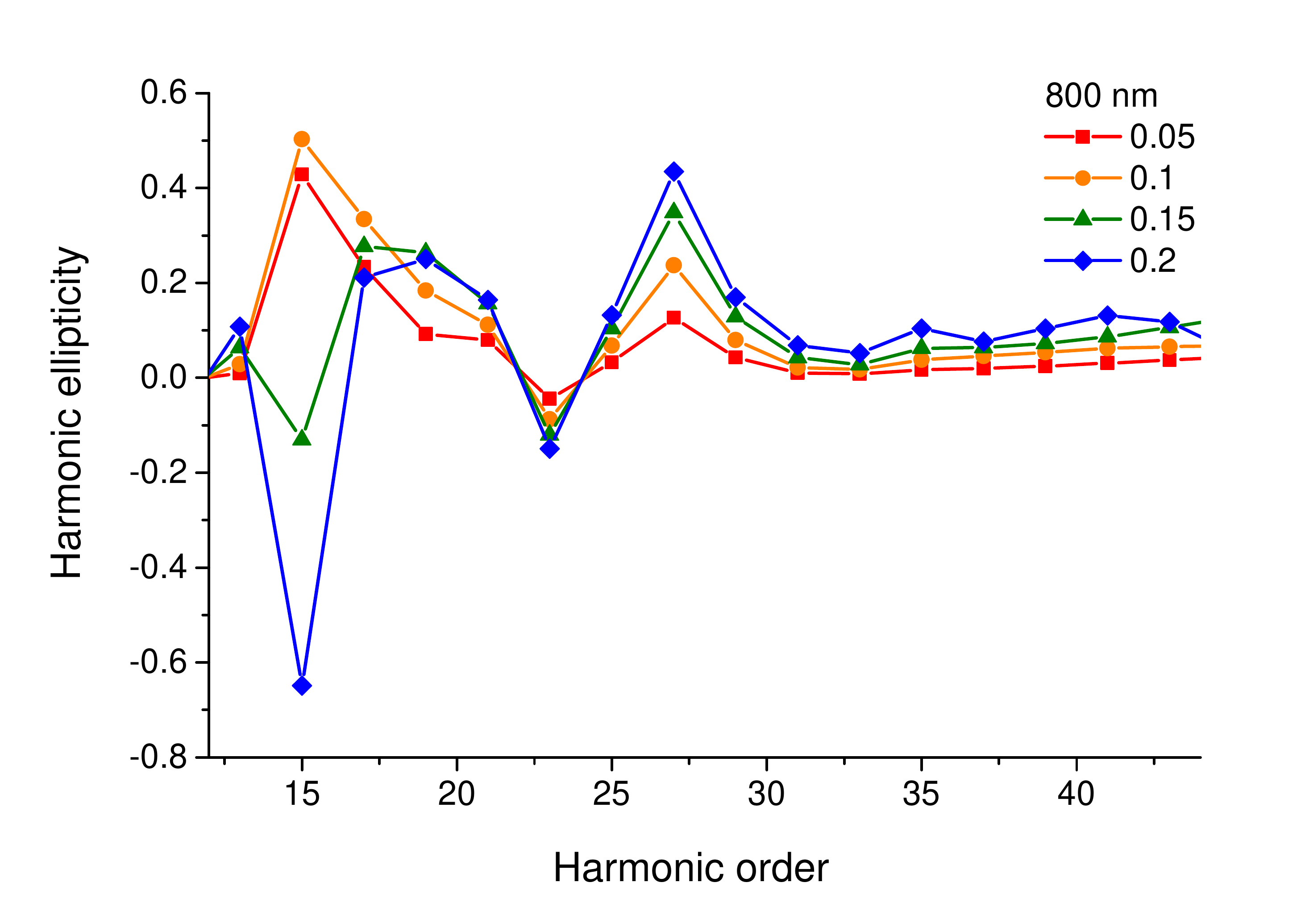}
\caption{Harmonic ellipticity as a function of the harmonic order for the fundamental wavelength 800~nm and fundamental ellipticities from 0.05 to 0.2 calculated within the SAE TDSE.}
\label{fig:ell_num}
\end{figure}

However, the most interesting and practically important characteristic of emitted harmonics is their ellipticity. It is calculated using \eqref{num_ell}, and presented in Fig.~\ref{fig:ell_num} for the 800~nm fundamental wavelength as a function of the harmonic order. Here it is shown that the harmonics around the resonance (H17) have higher ellipticity absolute values than nonresonant ones in the full absence of resonance~\cite{Ell_Origin,Strelkov_PRA_EPL}. It is important to note that the ellipticity of several harmonics above the resonance appears to be also affected by the presence of the resonance (even though H15, just below the resonance, has a rather high ellipticity, it is of limited interest due to its low intensity). This means that a number of harmonics around the resonance can be used for the creation of EP XUV short pulses. 

Note that there is one more region of high ellipticities in Fig.~\ref{fig:ell_num} lying around H27. This irregular behaviour looks like due to a Cooper minimum in the recombination cross section of the generating particle~\cite{Strelkov_PRA_EPL}. However, strictly speaking, in our calculations there cannot be any interference of recombination pathways as far as the ground state is $s$-orbital. The possible reason can be either quantum-path interference effects or an influence of the model potential used in our calculations, which is similar to the Ramsauer-Townsend effect, as well as resonances with the dressed AIS~\cite{Fareed2017}.

Figure~\ref{fig:ellipt} presents the behaviour of the ellipticity of resonant-nonresonant harmonic couples for each of wavelengths under consideration. Nonresonant harmonic ellipticities are presented up to the fundamental ellipticity, at which the harmonic yield drops by two orders of magnitude with respect to the yield at zero fundamental ellipticity. Their absolute values increase slowly with the fundamental ellipticity, as was shown in~\cite{Antoine_Lew,Strelkov_PRA_EPL}, for all considered wavelengths. 

\begin{figure}[t!]\centering 
\includegraphics[width=1.0\columnwidth]{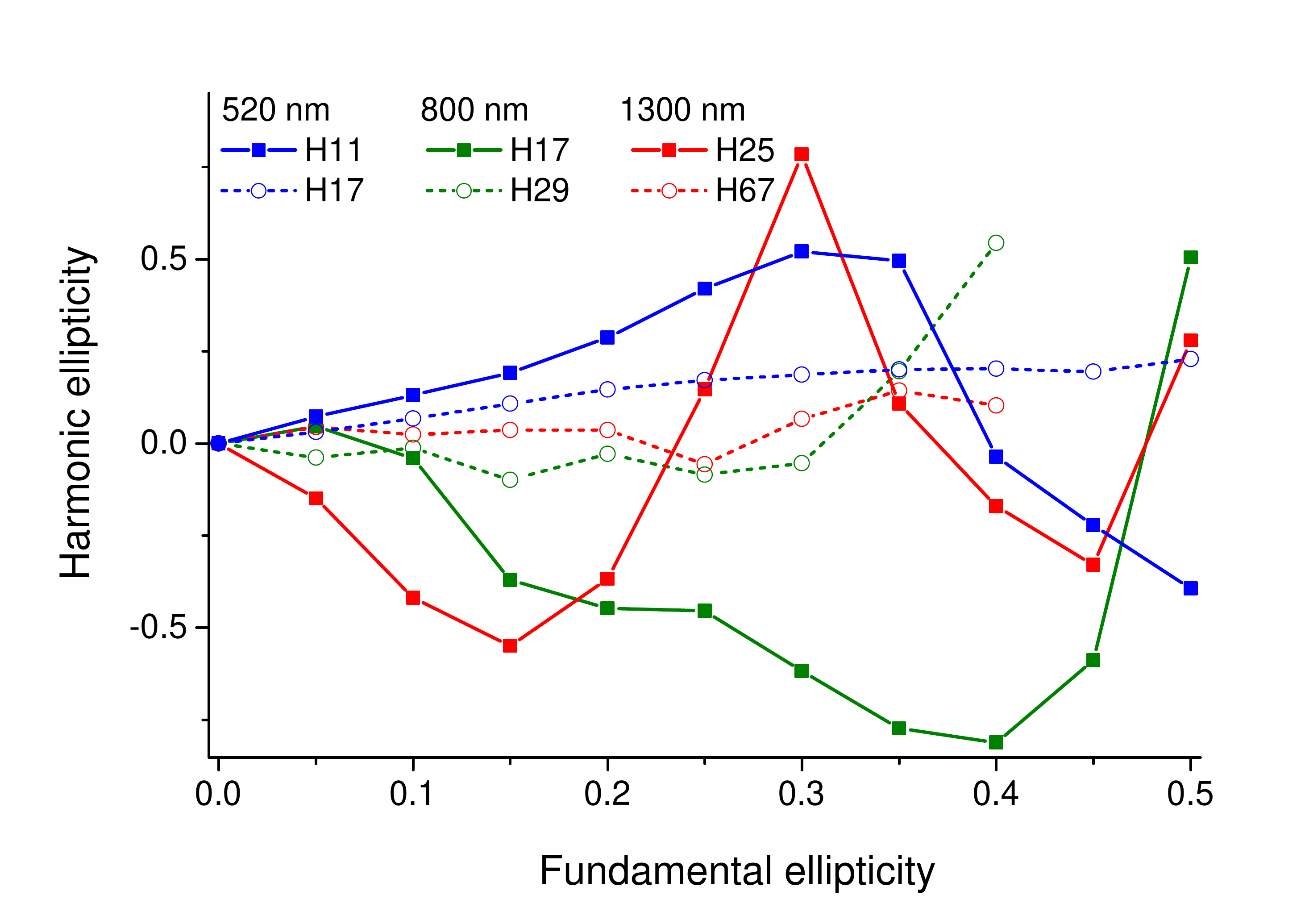}
\caption{Resonant (solid) and nonresonant (dashed) harmonic ellipticity as a function of the fundamental ellipticity calculated for the fundamental wavelengths 520~nm (blue squares), 800~nm (green circles) and 1300~nm (red triangles) within the SAE TDSE.}
\label{fig:ellipt}
\end{figure}

For the resonant harmonic ellipticities, on the contrary, the behaviour is much richer: the resonant harmonic ellipticity grows faster in absolute value than the nonresonant one for low fundamental ellipticities; moreover, its absolute value increases with the fundamental wavelength. Then the resonant harmonic ellipticity performs an anomalous behaviour: after reaching rather high harmonic ellipticity absolute values, it starts changing to high values of opposite sign. The maximum values of the resonant harmonic ellipticities presented in our calculations are 0.52 (at fundamental ellipticity 0.3), 0.81 (at fundamental ellipticity 0.4), and 0.78 (at fundamental ellipticity 0.3) for fundamental field wavelengths 520~nm, 800~nm, and 1300~nm, respectively.

Finally, it is important to note that the focus of this work is centred on properties of the microscopic harmonic response. An investigation of the macroscopic aspects of the problem, in particular, the coherence of this response, is a natural avenue for further studies. The promise directing us towards the study of macroscopic effects of HHG in EP pulses is provided by observations~\cite{Ganeev2014} showing that the coherences of resonant and nonresonant harmonics generated by a linearly-polarised driver in a laser plume are not only close to each other, but even, in principle, higher than the coherence of nonresonant harmonics generated in a gas jet.

\section{Conclusions and Outlook}
To summarise, in this study we conduct numerical integration of the 3D TDSE for resonant HHG in an EP laser field and also present a toy model of this process explaining our numerical results. Our calculations are carried out for three different wavelengths and show similar behaviour for all of them.

We show that the resonant harmonic yield behaves anomalously as opposed to the rapid decrease of the nonresonant harmonic with the fundamental ellipticity. As a result, for the ellipticities above the threshold one, only the resonant harmonic is generated. This can be explained within the toy model: as far as the AIS is much less localised than the ground state, the electron returning back to the parent particle is much more likely to be trapped into the AIS than directly recombine into the ground state in the EP field, especially for laser ellipticities larger than the threshold one.

Moreover, the ellipticity of resonant harmonics is higher than the nonresonant ones, without compromising the harmonic yield. This offers us a new route for the generation of quasi-monochromatic XUV with high ellipticity.

As a future direction of studies, we would suggest to search for better generating particles considering not only metallic ions but also the well-known giant resonances in rare gases~\cite{Shiner2011,Strelkov2016}, providing a better (wider or in a different spectral region) range of enhancement and higher yield, as well as to study the effect of the detuning from the resonance, which should provide a handle to control the ellipticity of the resonant harmonic, as well as nonresonat ones around it, and correspondingly, of the emitted XUV pulses.

\section*{Acknowledgments}
We acknowledge funding from RFBR [grant no. 19-02-00739] (M.K. and V.S.), the Foundation for the Advancement of Theoretical Physics and Mathematics ``BASIS'' (V.S.), and the Ministry of Science and Higher Education of the Russian Federation [state assignment for the Institute of Applied Physics RAS, project no. 0030-2021-0012] (M.E. and M.R.). 
M.K. acknowledges funding from the Alexander von Humboldt Foundation and support by Japan Science and Technology Agency (JST).
M.K. also acknowledges K.~Ishikawa and T.~Sato for sharing their many-body expertise.

\bibliography{lit}

\begin{thebibliography}{56}%
\makeatletter
\providecommand \@ifxundefined [1]{%
 \@ifx{#1\undefined}
}%
\providecommand \@ifnum [1]{%
 \ifnum #1\expandafter \@firstoftwo
 \else \expandafter \@secondoftwo
 \fi
}%
\providecommand \@ifx [1]{%
 \ifx #1\expandafter \@firstoftwo
 \else \expandafter \@secondoftwo
 \fi
}%
\providecommand \natexlab [1]{#1}%
\providecommand \enquote  [1]{``#1''}%
\providecommand \bibnamefont  [1]{#1}%
\providecommand \bibfnamefont [1]{#1}%
\providecommand \citenamefont [1]{#1}%
\providecommand \href@noop [0]{\@secondoftwo}%
\providecommand \href [0]{\begingroup \@sanitize@url \@href}%
\providecommand \@href[1]{\@@startlink{#1}\@@href}%
\providecommand \@@href[1]{\endgroup#1\@@endlink}%
\providecommand \@sanitize@url [0]{\catcode `\\12\catcode `\$12\catcode
  `\&12\catcode `\#12\catcode `\^12\catcode `\_12\catcode `\%12\relax}%
\providecommand \@@startlink[1]{}%
\providecommand \@@endlink[0]{}%
\providecommand \url  [0]{\begingroup\@sanitize@url \@url }%
\providecommand \@url [1]{\endgroup\@href {#1}{\urlprefix }}%
\providecommand \urlprefix  [0]{URL }%
\providecommand \Eprint [0]{\href }%
\providecommand \doibase [0]{https://doi.org/}%
\providecommand \selectlanguage [0]{\@gobble}%
\providecommand \bibinfo  [0]{\@secondoftwo}%
\providecommand \bibfield  [0]{\@secondoftwo}%
\providecommand \translation [1]{[#1]}%
\providecommand \BibitemOpen [0]{}%
\providecommand \bibitemStop [0]{}%
\providecommand \bibitemNoStop [0]{.\EOS\space}%
\providecommand \EOS [0]{\spacefactor3000\relax}%
\providecommand \BibitemShut  [1]{\csname bibitem#1\endcsname}%
\let\auto@bib@innerbib\@empty
\bibitem [{\citenamefont {Krausz}\ and\ \citenamefont
  {Ivanov}(2009)}]{atto_rev}%
  \BibitemOpen
  \bibfield  {author} {\bibinfo {author} {\bibfnamefont {F.}~\bibnamefont
  {Krausz}}\ and\ \bibinfo {author} {\bibfnamefont {M.}~\bibnamefont
  {Ivanov}},\ }\bibfield  {title} {\bibinfo {title} {Attosecond physics},\
  }\href {https://doi.org/10.1103/RevModPhys.81.163} {\bibfield  {journal}
  {\bibinfo  {journal} {Rev. Mod. Phys.}\ }\textbf {\bibinfo {volume} {81}},\
  \bibinfo {pages} {163} (\bibinfo {year} {2009})}\BibitemShut {NoStop}%
\bibitem [{\citenamefont {Pfeifer}\ \emph {et~al.}(2006)\citenamefont
  {Pfeifer}, \citenamefont {Spielmann},\ and\ \citenamefont
  {Gerber}}]{femto_rev}%
  \BibitemOpen
  \bibfield  {author} {\bibinfo {author} {\bibfnamefont {T.}~\bibnamefont
  {Pfeifer}}, \bibinfo {author} {\bibfnamefont {C.}~\bibnamefont {Spielmann}},\
  and\ \bibinfo {author} {\bibfnamefont {G.}~\bibnamefont {Gerber}},\
  }\bibfield  {title} {\bibinfo {title} {{Femtosecond x-ray science}},\ }\href
  {https://doi.org/10.1088/0034-4885/69/2/r04} {\bibfield  {journal} {\bibinfo
  {journal} {Rep. Prog. Phys.}\ }\textbf {\bibinfo {volume} {69}},\ \bibinfo
  {pages} {443} (\bibinfo {year} {2006})}\BibitemShut {NoStop}%
\bibitem [{\citenamefont {B\"owering}\ \emph {et~al.}(2001)\citenamefont
  {B\"owering}, \citenamefont {Lischke}, \citenamefont {Schmidtke},
  \citenamefont {M\"uller}, \citenamefont {Khalil},\ and\ \citenamefont
  {Heinzmann}}]{PhysRevLett.86.1187}%
  \BibitemOpen
  \bibfield  {author} {\bibinfo {author} {\bibfnamefont {N.}~\bibnamefont
  {B\"owering}}, \bibinfo {author} {\bibfnamefont {T.}~\bibnamefont {Lischke}},
  \bibinfo {author} {\bibfnamefont {B.}~\bibnamefont {Schmidtke}}, \bibinfo
  {author} {\bibfnamefont {N.}~\bibnamefont {M\"uller}}, \bibinfo {author}
  {\bibfnamefont {T.}~\bibnamefont {Khalil}},\ and\ \bibinfo {author}
  {\bibfnamefont {U.}~\bibnamefont {Heinzmann}},\ }\bibfield  {title} {\bibinfo
  {title} {Asymmetry in photoelectron emission from chiral molecules induced by
  circularly polarized light},\ }\href
  {https://doi.org/10.1103/PhysRevLett.86.1187} {\bibfield  {journal} {\bibinfo
   {journal} {Phys. Rev. Lett.}\ }\textbf {\bibinfo {volume} {86}},\ \bibinfo
  {pages} {1187} (\bibinfo {year} {2001})}\BibitemShut {NoStop}%
\bibitem [{\citenamefont {Beaulieu}\ \emph
  {et~al.}(2016{\natexlab{a}})\citenamefont {Beaulieu}, \citenamefont {Comby},
  \citenamefont {Fabre}, \citenamefont {Descamps}, \citenamefont
  {Ferr{\ifmmode\acute{e}\else\'{e}\fi}}, \citenamefont {Garcia}, \citenamefont
  {G{\ifmmode\acute{e}\else\'{e}\fi}neaux}, \citenamefont
  {L{\ifmmode\acute{e}\else\'{e}\fi}gar{\ifmmode\acute{e}\else\'{e}\fi}},
  \citenamefont {Nahon}, \citenamefont {Petit}, \citenamefont {Ruchon},
  \citenamefont {Pons}, \citenamefont {Blanchet},\ and\ \citenamefont
  {Mairesse}}]{Beaulieu2016Dec}%
  \BibitemOpen
  \bibfield  {author} {\bibinfo {author} {\bibfnamefont {S.}~\bibnamefont
  {Beaulieu}}, \bibinfo {author} {\bibfnamefont {A.}~\bibnamefont {Comby}},
  \bibinfo {author} {\bibfnamefont {B.}~\bibnamefont {Fabre}}, \bibinfo
  {author} {\bibfnamefont {D.}~\bibnamefont {Descamps}}, \bibinfo {author}
  {\bibfnamefont {A.}~\bibnamefont {Ferr{\ifmmode\acute{e}\else\'{e}\fi}}},
  \bibinfo {author} {\bibfnamefont {G.}~\bibnamefont {Garcia}}, \bibinfo
  {author} {\bibfnamefont {R.}~\bibnamefont
  {G{\ifmmode\acute{e}\else\'{e}\fi}neaux}}, \bibinfo {author} {\bibfnamefont
  {F.}~\bibnamefont
  {L{\ifmmode\acute{e}\else\'{e}\fi}gar{\ifmmode\acute{e}\else\'{e}\fi}}},
  \bibinfo {author} {\bibfnamefont {L.}~\bibnamefont {Nahon}}, \bibinfo
  {author} {\bibfnamefont {S.}~\bibnamefont {Petit}}, \bibinfo {author}
  {\bibfnamefont {T.}~\bibnamefont {Ruchon}}, \bibinfo {author} {\bibfnamefont
  {B.}~\bibnamefont {Pons}}, \bibinfo {author} {\bibfnamefont {V.}~\bibnamefont
  {Blanchet}},\ and\ \bibinfo {author} {\bibfnamefont {Y.}~\bibnamefont
  {Mairesse}},\ }\bibfield  {title} {\bibinfo {title} {{Probing ultrafast
  dynamics of chiral molecules using time-resolved photoelectron circular
  dichroism}},\ }\href {https://doi.org/10.1039/C6FD00113K} {\bibfield
  {journal} {\bibinfo  {journal} {Faraday Discuss.}\ }\textbf {\bibinfo
  {volume} {194}},\ \bibinfo {pages} {325} (\bibinfo {year}
  {2016}{\natexlab{a}})}\BibitemShut {NoStop}%
\bibitem [{\citenamefont {Boeglin}\ \emph {et~al.}(2010)\citenamefont
  {Boeglin}, \citenamefont {Beaurepaire}, \citenamefont
  {Halt{\ifmmode\acute{e}\else\'{e}\fi}}, \citenamefont
  {L{\ifmmode\acute{o}\else\'{o}\fi}pez-Flores}, \citenamefont {Stamm},
  \citenamefont {Pontius}, \citenamefont {D{\ifmmode\ddot{u}\else\"{u}\fi}rr},\
  and\ \citenamefont {Bigot}}]{Boeglin2010May}%
  \BibitemOpen
  \bibfield  {author} {\bibinfo {author} {\bibfnamefont {C.}~\bibnamefont
  {Boeglin}}, \bibinfo {author} {\bibfnamefont {E.}~\bibnamefont
  {Beaurepaire}}, \bibinfo {author} {\bibfnamefont {V.}~\bibnamefont
  {Halt{\ifmmode\acute{e}\else\'{e}\fi}}}, \bibinfo {author} {\bibfnamefont
  {V.}~\bibnamefont {L{\ifmmode\acute{o}\else\'{o}\fi}pez-Flores}}, \bibinfo
  {author} {\bibfnamefont {C.}~\bibnamefont {Stamm}}, \bibinfo {author}
  {\bibfnamefont {N.}~\bibnamefont {Pontius}}, \bibinfo {author} {\bibfnamefont
  {H.~A.}\ \bibnamefont {D{\ifmmode\ddot{u}\else\"{u}\fi}rr}},\ and\ \bibinfo
  {author} {\bibfnamefont {J.-Y.}\ \bibnamefont {Bigot}},\ }\bibfield  {title}
  {\bibinfo {title} {{Distinguishing the ultrafast dynamics of spin and orbital
  moments in solids}},\ }\href {https://doi.org/10.1038/nature09070} {\bibfield
   {journal} {\bibinfo  {journal} {Nature}\ }\textbf {\bibinfo {volume}
  {465}},\ \bibinfo {pages} {458} (\bibinfo {year} {2010})}\BibitemShut
  {NoStop}%
\bibitem [{\citenamefont {Kfir}\ \emph {et~al.}(2017)\citenamefont {Kfir},
  \citenamefont {Zayko}, \citenamefont {Nolte}, \citenamefont {Sivis},
  \citenamefont {M{\ifmmode\ddot{o}\else\"{o}\fi}ller}, \citenamefont {Hebler},
  \citenamefont {Arekapudi}, \citenamefont {Steil}, \citenamefont
  {Sch{\ifmmode\ddot{a}\else\"{a}\fi}fer}, \citenamefont {Albrecht},
  \citenamefont {Cohen}, \citenamefont {Mathias},\ and\ \citenamefont
  {Ropers}}]{Kfir2017Dec}%
  \BibitemOpen
  \bibfield  {author} {\bibinfo {author} {\bibfnamefont {O.}~\bibnamefont
  {Kfir}}, \bibinfo {author} {\bibfnamefont {S.}~\bibnamefont {Zayko}},
  \bibinfo {author} {\bibfnamefont {C.}~\bibnamefont {Nolte}}, \bibinfo
  {author} {\bibfnamefont {M.}~\bibnamefont {Sivis}}, \bibinfo {author}
  {\bibfnamefont {M.}~\bibnamefont {M{\ifmmode\ddot{o}\else\"{o}\fi}ller}},
  \bibinfo {author} {\bibfnamefont {B.}~\bibnamefont {Hebler}}, \bibinfo
  {author} {\bibfnamefont {S.~S. P.~K.}\ \bibnamefont {Arekapudi}}, \bibinfo
  {author} {\bibfnamefont {D.}~\bibnamefont {Steil}}, \bibinfo {author}
  {\bibfnamefont {S.}~\bibnamefont {Sch{\ifmmode\ddot{a}\else\"{a}\fi}fer}},
  \bibinfo {author} {\bibfnamefont {M.}~\bibnamefont {Albrecht}}, \bibinfo
  {author} {\bibfnamefont {O.}~\bibnamefont {Cohen}}, \bibinfo {author}
  {\bibfnamefont {S.}~\bibnamefont {Mathias}},\ and\ \bibinfo {author}
  {\bibfnamefont {C.}~\bibnamefont {Ropers}},\ }\bibfield  {title} {\bibinfo
  {title} {{Nanoscale magnetic imaging using circularly polarized high-harmonic
  radiation}},\ }\href {https://doi.org/10.1126/sciadv.aao4641} {\bibfield
  {journal} {\bibinfo  {journal} {Sci. Adv.}\ }\textbf {\bibinfo {volume}
  {3}},\ \bibinfo {pages} {eaao4641} (\bibinfo {year} {2017})}\BibitemShut
  {NoStop}%
\bibitem [{\citenamefont {Willems}\ \emph {et~al.}(2015)\citenamefont
  {Willems}, \citenamefont {Smeenk}, \citenamefont {Zhavoronkov}, \citenamefont
  {Kornilov}, \citenamefont {Radu}, \citenamefont {Schmidbauer}, \citenamefont
  {Hanke}, \citenamefont {von Korff~Schmising}, \citenamefont {Vrakking},\ and\
  \citenamefont {Eisebitt}}]{PhysRevB.92.220405}%
  \BibitemOpen
  \bibfield  {author} {\bibinfo {author} {\bibfnamefont {F.}~\bibnamefont
  {Willems}}, \bibinfo {author} {\bibfnamefont {C.~T.~L.}\ \bibnamefont
  {Smeenk}}, \bibinfo {author} {\bibfnamefont {N.}~\bibnamefont {Zhavoronkov}},
  \bibinfo {author} {\bibfnamefont {O.}~\bibnamefont {Kornilov}}, \bibinfo
  {author} {\bibfnamefont {I.}~\bibnamefont {Radu}}, \bibinfo {author}
  {\bibfnamefont {M.}~\bibnamefont {Schmidbauer}}, \bibinfo {author}
  {\bibfnamefont {M.}~\bibnamefont {Hanke}}, \bibinfo {author} {\bibfnamefont
  {C.}~\bibnamefont {von Korff~Schmising}}, \bibinfo {author} {\bibfnamefont
  {M.~J.~J.}\ \bibnamefont {Vrakking}},\ and\ \bibinfo {author} {\bibfnamefont
  {S.}~\bibnamefont {Eisebitt}},\ }\bibfield  {title} {\bibinfo {title}
  {Probing ultrafast spin dynamics with high-harmonic magnetic circular
  dichroism spectroscopy},\ }\href {https://doi.org/10.1103/PhysRevB.92.220405}
  {\bibfield  {journal} {\bibinfo  {journal} {Phys. Rev. B}\ }\textbf {\bibinfo
  {volume} {92}},\ \bibinfo {pages} {220405(R)} (\bibinfo {year}
  {2015})}\BibitemShut {NoStop}%
\bibitem [{\citenamefont {Siegrist}\ \emph {et~al.}(2019)\citenamefont
  {Siegrist}, \citenamefont {Gessner}, \citenamefont {Ossiander}, \citenamefont
  {Denker}, \citenamefont {Chang}, \citenamefont
  {Schr{\ifmmode\ddot{o}\else\"{o}\fi}der}, \citenamefont {Guggenmos},
  \citenamefont {Cui}, \citenamefont {Walowski}, \citenamefont {Martens},
  \citenamefont {Dewhurst}, \citenamefont {Kleineberg}, \citenamefont
  {M{\ifmmode\ddot{u}\else\"{u}\fi}nzenberg}, \citenamefont {Sharma},\ and\
  \citenamefont {Schultze}}]{Siegrist2019Jun}%
  \BibitemOpen
  \bibfield  {author} {\bibinfo {author} {\bibfnamefont {F.}~\bibnamefont
  {Siegrist}}, \bibinfo {author} {\bibfnamefont {J.~A.}\ \bibnamefont
  {Gessner}}, \bibinfo {author} {\bibfnamefont {M.}~\bibnamefont {Ossiander}},
  \bibinfo {author} {\bibfnamefont {C.}~\bibnamefont {Denker}}, \bibinfo
  {author} {\bibfnamefont {Y.-P.}\ \bibnamefont {Chang}}, \bibinfo {author}
  {\bibfnamefont {M.~C.}\ \bibnamefont
  {Schr{\ifmmode\ddot{o}\else\"{o}\fi}der}}, \bibinfo {author} {\bibfnamefont
  {A.}~\bibnamefont {Guggenmos}}, \bibinfo {author} {\bibfnamefont
  {Y.}~\bibnamefont {Cui}}, \bibinfo {author} {\bibfnamefont {J.}~\bibnamefont
  {Walowski}}, \bibinfo {author} {\bibfnamefont {U.}~\bibnamefont {Martens}},
  \bibinfo {author} {\bibfnamefont {J.~K.}\ \bibnamefont {Dewhurst}}, \bibinfo
  {author} {\bibfnamefont {U.}~\bibnamefont {Kleineberg}}, \bibinfo {author}
  {\bibfnamefont {M.}~\bibnamefont {M{\ifmmode\ddot{u}\else\"{u}\fi}nzenberg}},
  \bibinfo {author} {\bibfnamefont {S.}~\bibnamefont {Sharma}},\ and\ \bibinfo
  {author} {\bibfnamefont {M.}~\bibnamefont {Schultze}},\ }\bibfield  {title}
  {\bibinfo {title} {{Light-wave dynamic control of magnetism}},\ }\href
  {https://doi.org/10.1038/s41586-019-1333-x} {\bibfield  {journal} {\bibinfo
  {journal} {Nature}\ }\textbf {\bibinfo {volume} {571}},\ \bibinfo {pages}
  {240} (\bibinfo {year} {2019})}\BibitemShut {NoStop}%
\bibitem [{\citenamefont {Donsa}\ \emph {et~al.}(2019)\citenamefont {Donsa},
  \citenamefont {B\ifmmode~\check{r}\else \v{r}\fi{}ezinov\'a}, \citenamefont
  {Ni}, \citenamefont {Feist},\ and\ \citenamefont
  {Burgd\"orfer}}]{PhysRevA.99.023413}%
  \BibitemOpen
  \bibfield  {author} {\bibinfo {author} {\bibfnamefont {S.}~\bibnamefont
  {Donsa}}, \bibinfo {author} {\bibfnamefont {I.}~\bibnamefont
  {B\ifmmode~\check{r}\else \v{r}\fi{}ezinov\'a}}, \bibinfo {author}
  {\bibfnamefont {H.}~\bibnamefont {Ni}}, \bibinfo {author} {\bibfnamefont
  {J.}~\bibnamefont {Feist}},\ and\ \bibinfo {author} {\bibfnamefont
  {J.}~\bibnamefont {Burgd\"orfer}},\ }\bibfield  {title} {\bibinfo {title}
  {Polarization tagging of two-photon double ionization by elliptically
  polarized {XUV} pulses},\ }\href {https://doi.org/10.1103/PhysRevA.99.023413}
  {\bibfield  {journal} {\bibinfo  {journal} {Phys. Rev. A}\ }\textbf {\bibinfo
  {volume} {99}},\ \bibinfo {pages} {023413} (\bibinfo {year}
  {2019})}\BibitemShut {NoStop}%
\bibitem [{\citenamefont {Prost}\ \emph {et~al.}(2018)\citenamefont {Prost},
  \citenamefont {Hertz}, \citenamefont {Billard}, \citenamefont {Lavorel},\
  and\ \citenamefont {Faucher}}]{Prost2018Nov}%
  \BibitemOpen
  \bibfield  {author} {\bibinfo {author} {\bibfnamefont {E.}~\bibnamefont
  {Prost}}, \bibinfo {author} {\bibfnamefont {E.}~\bibnamefont {Hertz}},
  \bibinfo {author} {\bibfnamefont {F.}~\bibnamefont {Billard}}, \bibinfo
  {author} {\bibfnamefont {B.}~\bibnamefont {Lavorel}},\ and\ \bibinfo {author}
  {\bibfnamefont {O.}~\bibnamefont {Faucher}},\ }\bibfield  {title} {\bibinfo
  {title} {{Polarization-based tachometer for measuring spinning rotors}},\
  }\href {https://doi.org/10.1364/OE.26.031839} {\bibfield  {journal} {\bibinfo
   {journal} {Opt. Express}\ }\textbf {\bibinfo {volume} {26}},\ \bibinfo
  {pages} {31839} (\bibinfo {year} {2018})}\BibitemShut {NoStop}%
\bibitem [{\citenamefont {Allaria}\ \emph {et~al.}(2014)\citenamefont
  {Allaria}, \citenamefont {Diviacco}, \citenamefont {Callegari}, \citenamefont
  {Finetti}, \citenamefont {Mahieu}, \citenamefont {Viefhaus}, \citenamefont
  {Zangrando}, \citenamefont {De~Ninno}, \citenamefont {Lambert}, \citenamefont
  {Ferrari}, \citenamefont {Buck}, \citenamefont {Ilchen}, \citenamefont
  {Vodungbo}, \citenamefont {Mahne}, \citenamefont {Svetina}, \citenamefont
  {Spezzani}, \citenamefont {Di~Mitri}, \citenamefont {Penco}, \citenamefont
  {Trov\'o}, \citenamefont {Fawley}, \citenamefont {Rebernik}, \citenamefont
  {Gauthier}, \citenamefont {Grazioli}, \citenamefont {Coreno}, \citenamefont
  {Ressel}, \citenamefont {Kivim\"aki}, \citenamefont {Mazza}, \citenamefont
  {Glaser}, \citenamefont {Scholz}, \citenamefont {Seltmann}, \citenamefont
  {Gessler}, \citenamefont {Gr\"unert}, \citenamefont {De~Fanis}, \citenamefont
  {Meyer}, \citenamefont {Knie}, \citenamefont {Moeller}, \citenamefont
  {Raimondi}, \citenamefont {Capotondi}, \citenamefont {Pedersoli},
  \citenamefont {Plekan}, \citenamefont {Danailov}, \citenamefont {Demidovich},
  \citenamefont {Nikolov}, \citenamefont {Abrami}, \citenamefont {Gautier},
  \citenamefont {L\"uning}, \citenamefont {Zeitoun},\ and\ \citenamefont
  {Giannessi}}]{PhysRevX.4.041040}%
  \BibitemOpen
  \bibfield  {author} {\bibinfo {author} {\bibfnamefont {E.}~\bibnamefont
  {Allaria}}, \bibinfo {author} {\bibfnamefont {B.}~\bibnamefont {Diviacco}},
  \bibinfo {author} {\bibfnamefont {C.}~\bibnamefont {Callegari}}, \bibinfo
  {author} {\bibfnamefont {P.}~\bibnamefont {Finetti}}, \bibinfo {author}
  {\bibfnamefont {B.}~\bibnamefont {Mahieu}}, \bibinfo {author} {\bibfnamefont
  {J.}~\bibnamefont {Viefhaus}}, \bibinfo {author} {\bibfnamefont
  {M.}~\bibnamefont {Zangrando}}, \bibinfo {author} {\bibfnamefont
  {G.}~\bibnamefont {De~Ninno}}, \bibinfo {author} {\bibfnamefont
  {G.}~\bibnamefont {Lambert}}, \bibinfo {author} {\bibfnamefont
  {E.}~\bibnamefont {Ferrari}}, \bibinfo {author} {\bibfnamefont
  {J.}~\bibnamefont {Buck}}, \bibinfo {author} {\bibfnamefont {M.}~\bibnamefont
  {Ilchen}}, \bibinfo {author} {\bibfnamefont {B.}~\bibnamefont {Vodungbo}},
  \bibinfo {author} {\bibfnamefont {N.}~\bibnamefont {Mahne}}, \bibinfo
  {author} {\bibfnamefont {C.}~\bibnamefont {Svetina}}, \bibinfo {author}
  {\bibfnamefont {C.}~\bibnamefont {Spezzani}}, \bibinfo {author}
  {\bibfnamefont {S.}~\bibnamefont {Di~Mitri}}, \bibinfo {author}
  {\bibfnamefont {G.}~\bibnamefont {Penco}}, \bibinfo {author} {\bibfnamefont
  {M.}~\bibnamefont {Trov\'o}}, \bibinfo {author} {\bibfnamefont {W.~M.}\
  \bibnamefont {Fawley}}, \bibinfo {author} {\bibfnamefont {P.~R.}\
  \bibnamefont {Rebernik}}, \bibinfo {author} {\bibfnamefont {D.}~\bibnamefont
  {Gauthier}}, \bibinfo {author} {\bibfnamefont {C.}~\bibnamefont {Grazioli}},
  \bibinfo {author} {\bibfnamefont {M.}~\bibnamefont {Coreno}}, \bibinfo
  {author} {\bibfnamefont {B.}~\bibnamefont {Ressel}}, \bibinfo {author}
  {\bibfnamefont {A.}~\bibnamefont {Kivim\"aki}}, \bibinfo {author}
  {\bibfnamefont {T.}~\bibnamefont {Mazza}}, \bibinfo {author} {\bibfnamefont
  {L.}~\bibnamefont {Glaser}}, \bibinfo {author} {\bibfnamefont
  {F.}~\bibnamefont {Scholz}}, \bibinfo {author} {\bibfnamefont
  {J.}~\bibnamefont {Seltmann}}, \bibinfo {author} {\bibfnamefont
  {P.}~\bibnamefont {Gessler}}, \bibinfo {author} {\bibfnamefont
  {J.}~\bibnamefont {Gr\"unert}}, \bibinfo {author} {\bibfnamefont
  {A.}~\bibnamefont {De~Fanis}}, \bibinfo {author} {\bibfnamefont
  {M.}~\bibnamefont {Meyer}}, \bibinfo {author} {\bibfnamefont
  {A.}~\bibnamefont {Knie}}, \bibinfo {author} {\bibfnamefont {S.~P.}\
  \bibnamefont {Moeller}}, \bibinfo {author} {\bibfnamefont {L.}~\bibnamefont
  {Raimondi}}, \bibinfo {author} {\bibfnamefont {F.}~\bibnamefont {Capotondi}},
  \bibinfo {author} {\bibfnamefont {E.}~\bibnamefont {Pedersoli}}, \bibinfo
  {author} {\bibfnamefont {O.}~\bibnamefont {Plekan}}, \bibinfo {author}
  {\bibfnamefont {M.~B.}\ \bibnamefont {Danailov}}, \bibinfo {author}
  {\bibfnamefont {A.}~\bibnamefont {Demidovich}}, \bibinfo {author}
  {\bibfnamefont {I.}~\bibnamefont {Nikolov}}, \bibinfo {author} {\bibfnamefont
  {A.}~\bibnamefont {Abrami}}, \bibinfo {author} {\bibfnamefont
  {J.}~\bibnamefont {Gautier}}, \bibinfo {author} {\bibfnamefont
  {J.}~\bibnamefont {L\"uning}}, \bibinfo {author} {\bibfnamefont
  {P.}~\bibnamefont {Zeitoun}},\ and\ \bibinfo {author} {\bibfnamefont
  {L.}~\bibnamefont {Giannessi}},\ }\bibfield  {title} {\bibinfo {title}
  {Control of the polarization of a vacuum-ultraviolet, high-gain,
  free-electron laser},\ }\href {https://doi.org/10.1103/PhysRevX.4.041040}
  {\bibfield  {journal} {\bibinfo  {journal} {Phys. Rev. X}\ }\textbf {\bibinfo
  {volume} {4}},\ \bibinfo {pages} {041040} (\bibinfo {year}
  {2014})}\BibitemShut {NoStop}%
\bibitem [{\citenamefont {Budil}\ \emph {et~al.}(1993)\citenamefont {Budil},
  \citenamefont {Sali\`eres}, \citenamefont {L'Huillier}, \citenamefont
  {Ditmire},\ and\ \citenamefont {Perry}}]{Budil_PRA_48}%
  \BibitemOpen
  \bibfield  {author} {\bibinfo {author} {\bibfnamefont {K.~S.}\ \bibnamefont
  {Budil}}, \bibinfo {author} {\bibfnamefont {P.}~\bibnamefont {Sali\`eres}},
  \bibinfo {author} {\bibfnamefont {A.}~\bibnamefont {L'Huillier}}, \bibinfo
  {author} {\bibfnamefont {T.}~\bibnamefont {Ditmire}},\ and\ \bibinfo {author}
  {\bibfnamefont {M.~D.}\ \bibnamefont {Perry}},\ }\bibfield  {title} {\bibinfo
  {title} {Influence of ellipticity on harmonic generation},\ }\href
  {https://doi.org/10.1103/PhysRevA.48.R3437} {\bibfield  {journal} {\bibinfo
  {journal} {Phys. Rev. A}\ }\textbf {\bibinfo {volume} {48}},\ \bibinfo
  {pages} {R3437} (\bibinfo {year} {1993})}\BibitemShut {NoStop}%
\bibitem [{\citenamefont {Antoine}\ \emph {et~al.}(1997)\citenamefont
  {Antoine}, \citenamefont {Carr\'e}, \citenamefont {L'Huillier},\ and\
  \citenamefont {Lewenstein}}]{Antoine_Lew}%
  \BibitemOpen
  \bibfield  {author} {\bibinfo {author} {\bibfnamefont {P.}~\bibnamefont
  {Antoine}}, \bibinfo {author} {\bibfnamefont {B.}~\bibnamefont {Carr\'e}},
  \bibinfo {author} {\bibfnamefont {A.}~\bibnamefont {L'Huillier}},\ and\
  \bibinfo {author} {\bibfnamefont {M.}~\bibnamefont {Lewenstein}},\ }\bibfield
   {title} {\bibinfo {title} {Polarization of high-order harmonics},\ }\href
  {https://doi.org/10.1103/PhysRevA.55.1314} {\bibfield  {journal} {\bibinfo
  {journal} {Phys. Rev. A}\ }\textbf {\bibinfo {volume} {55}},\ \bibinfo
  {pages} {1314} (\bibinfo {year} {1997})}\BibitemShut {NoStop}%
\bibitem [{\citenamefont {Dietrich}\ \emph {et~al.}(1994)\citenamefont
  {Dietrich}, \citenamefont {Burnett}, \citenamefont {Ivanov},\ and\
  \citenamefont {Corkum}}]{Dietrich1994}%
  \BibitemOpen
  \bibfield  {author} {\bibinfo {author} {\bibfnamefont {P.}~\bibnamefont
  {Dietrich}}, \bibinfo {author} {\bibfnamefont {N.~H.}\ \bibnamefont
  {Burnett}}, \bibinfo {author} {\bibfnamefont {M.}~\bibnamefont {Ivanov}},\
  and\ \bibinfo {author} {\bibfnamefont {P.~B.}\ \bibnamefont {Corkum}},\
  }\bibfield  {title} {\bibinfo {title} {{High-harmonic generation and
  correlated two-electron multiphoton ionization with elliptically polarized
  light}},\ }\href {https://doi.org/10.1103/PhysRevA.50.R3585} {\bibfield
  {journal} {\bibinfo  {journal} {Phys. Rev. A}\ }\textbf {\bibinfo {volume}
  {50}},\ \bibinfo {pages} {R3585} (\bibinfo {year} {1994})}\BibitemShut
  {NoStop}%
\bibitem [{\citenamefont {M\"oller}\ \emph {et~al.}(2012)\citenamefont
  {M\"oller}, \citenamefont {Cheng}, \citenamefont {Khan}, \citenamefont
  {Zhao}, \citenamefont {Zhao}, \citenamefont {Chini}, \citenamefont {Paulus},\
  and\ \citenamefont {Chang}}]{PhysRevA.86.011401}%
  \BibitemOpen
  \bibfield  {author} {\bibinfo {author} {\bibfnamefont {M.}~\bibnamefont
  {M\"oller}}, \bibinfo {author} {\bibfnamefont {Y.}~\bibnamefont {Cheng}},
  \bibinfo {author} {\bibfnamefont {S.~D.}\ \bibnamefont {Khan}}, \bibinfo
  {author} {\bibfnamefont {B.}~\bibnamefont {Zhao}}, \bibinfo {author}
  {\bibfnamefont {K.}~\bibnamefont {Zhao}}, \bibinfo {author} {\bibfnamefont
  {M.}~\bibnamefont {Chini}}, \bibinfo {author} {\bibfnamefont {G.~G.}\
  \bibnamefont {Paulus}},\ and\ \bibinfo {author} {\bibfnamefont
  {Z.}~\bibnamefont {Chang}},\ }\bibfield  {title} {\bibinfo {title}
  {Dependence of high-order-harmonic-generation yield on driving-laser
  ellipticity},\ }\href {https://doi.org/10.1103/PhysRevA.86.011401} {\bibfield
   {journal} {\bibinfo  {journal} {Phys. Rev. A}\ }\textbf {\bibinfo {volume}
  {86}},\ \bibinfo {pages} {011401(R)} (\bibinfo {year} {2012})}\BibitemShut
  {NoStop}%
\bibitem [{\citenamefont {Eichmann}\ \emph {et~al.}(1995)\citenamefont
  {Eichmann}, \citenamefont {Egbert}, \citenamefont {Nolte}, \citenamefont
  {Momma}, \citenamefont {Wellegehausen}, \citenamefont {Becker}, \citenamefont
  {Long},\ and\ \citenamefont {McIver}}]{Eichmann1995}%
  \BibitemOpen
  \bibfield  {author} {\bibinfo {author} {\bibfnamefont {H.}~\bibnamefont
  {Eichmann}}, \bibinfo {author} {\bibfnamefont {A.}~\bibnamefont {Egbert}},
  \bibinfo {author} {\bibfnamefont {S.}~\bibnamefont {Nolte}}, \bibinfo
  {author} {\bibfnamefont {C.}~\bibnamefont {Momma}}, \bibinfo {author}
  {\bibfnamefont {B.}~\bibnamefont {Wellegehausen}}, \bibinfo {author}
  {\bibfnamefont {W.}~\bibnamefont {Becker}}, \bibinfo {author} {\bibfnamefont
  {S.}~\bibnamefont {Long}},\ and\ \bibinfo {author} {\bibfnamefont {J.~K.}\
  \bibnamefont {McIver}},\ }\bibfield  {title} {\bibinfo {title}
  {{Polarization-dependent high-order two-color mixing}},\ }\href
  {https://doi.org/10.1103/PhysRevA.51.R3414} {\bibfield  {journal} {\bibinfo
  {journal} {Phys. Rev. A}\ }\textbf {\bibinfo {volume} {51}},\ \bibinfo
  {pages} {R3414} (\bibinfo {year} {1995})}\BibitemShut {NoStop}%
\bibitem [{\citenamefont {Long}\ \emph {et~al.}(1995)\citenamefont {Long},
  \citenamefont {Becker},\ and\ \citenamefont {McIver}}]{PhysRevA.52.2262}%
  \BibitemOpen
  \bibfield  {author} {\bibinfo {author} {\bibfnamefont {S.}~\bibnamefont
  {Long}}, \bibinfo {author} {\bibfnamefont {W.}~\bibnamefont {Becker}},\ and\
  \bibinfo {author} {\bibfnamefont {J.~K.}\ \bibnamefont {McIver}},\ }\bibfield
   {title} {\bibinfo {title} {Model calculations of polarization-dependent
  two-color high-harmonic generation},\ }\href
  {https://doi.org/10.1103/PhysRevA.52.2262} {\bibfield  {journal} {\bibinfo
  {journal} {Phys. Rev. A}\ }\textbf {\bibinfo {volume} {52}},\ \bibinfo
  {pages} {2262} (\bibinfo {year} {1995})}\BibitemShut {NoStop}%
\bibitem [{\citenamefont {Ivanov}\ and\ \citenamefont
  {Pisanty}(2014)}]{Emilio_Misha}%
  \BibitemOpen
  \bibfield  {author} {\bibinfo {author} {\bibfnamefont {M.}~\bibnamefont
  {Ivanov}}\ and\ \bibinfo {author} {\bibfnamefont {E.}~\bibnamefont
  {Pisanty}},\ }\bibfield  {title} {\bibinfo {title} {Taking control of
  polarization},\ }\href {https://doi.org/10.1038/nphoton.2014.141} {\bibfield
  {journal} {\bibinfo  {journal} {Nature Photonics}\ }\textbf {\bibinfo
  {volume} {8}},\ \bibinfo {pages} {501} (\bibinfo {year} {2014})}\BibitemShut
  {NoStop}%
\bibitem [{\citenamefont {Fleischer}\ \emph {et~al.}(2014)\citenamefont
  {Fleischer}, \citenamefont {Kfir}, \citenamefont {Diskin}, \citenamefont
  {Sidorenko},\ and\ \citenamefont {Cohen}}]{Fleischer:14}%
  \BibitemOpen
  \bibfield  {author} {\bibinfo {author} {\bibfnamefont {A.}~\bibnamefont
  {Fleischer}}, \bibinfo {author} {\bibfnamefont {O.}~\bibnamefont {Kfir}},
  \bibinfo {author} {\bibfnamefont {T.}~\bibnamefont {Diskin}}, \bibinfo
  {author} {\bibfnamefont {P.}~\bibnamefont {Sidorenko}},\ and\ \bibinfo
  {author} {\bibfnamefont {O.}~\bibnamefont {Cohen}},\ }\bibfield  {title}
  {\bibinfo {title} {Spin angular momentum and tunable polarization in
  high-harmonic generation},\ }\href {https://doi.org/10.1038/nphoton.2014.108}
  {\bibfield  {journal} {\bibinfo  {journal} {Nature Photonics}\ }\textbf
  {\bibinfo {volume} {8}},\ \bibinfo {pages} {543} (\bibinfo {year}
  {2014})}\BibitemShut {NoStop}%
\bibitem [{\citenamefont {Kfir}\ \emph {et~al.}(2015)\citenamefont {Kfir},
  \citenamefont {Grychtol}, \citenamefont {Turgut}, \citenamefont {Knut},
  \citenamefont {Zusin}, \citenamefont {Popmintchev}, \citenamefont
  {Popmintchev}, \citenamefont {Nembach}, \citenamefont {Shaw}, \citenamefont
  {Fleischer}, \citenamefont {Kapteyn}, \citenamefont {Murnane},\ and\
  \citenamefont {Cohen}}]{Kfir2014Dec}%
  \BibitemOpen
  \bibfield  {author} {\bibinfo {author} {\bibfnamefont {O.}~\bibnamefont
  {Kfir}}, \bibinfo {author} {\bibfnamefont {P.}~\bibnamefont {Grychtol}},
  \bibinfo {author} {\bibfnamefont {E.}~\bibnamefont {Turgut}}, \bibinfo
  {author} {\bibfnamefont {R.}~\bibnamefont {Knut}}, \bibinfo {author}
  {\bibfnamefont {D.}~\bibnamefont {Zusin}}, \bibinfo {author} {\bibfnamefont
  {D.}~\bibnamefont {Popmintchev}}, \bibinfo {author} {\bibfnamefont
  {T.}~\bibnamefont {Popmintchev}}, \bibinfo {author} {\bibfnamefont
  {H.}~\bibnamefont {Nembach}}, \bibinfo {author} {\bibfnamefont {J.~M.}\
  \bibnamefont {Shaw}}, \bibinfo {author} {\bibfnamefont {A.}~\bibnamefont
  {Fleischer}}, \bibinfo {author} {\bibfnamefont {H.}~\bibnamefont {Kapteyn}},
  \bibinfo {author} {\bibfnamefont {M.}~\bibnamefont {Murnane}},\ and\ \bibinfo
  {author} {\bibfnamefont {O.}~\bibnamefont {Cohen}},\ }\bibfield  {title}
  {\bibinfo {title} {{Generation of bright phase-matched circularly-polarized
  extreme ultraviolet high harmonics}},\ }\href
  {https://doi.org/10.1038/nphoton.2014.293} {\bibfield  {journal} {\bibinfo
  {journal} {Nat. Photonics}\ }\textbf {\bibinfo {volume} {9}},\ \bibinfo
  {pages} {99} (\bibinfo {year} {2015})}\BibitemShut {NoStop}%
\bibitem [{\citenamefont {Fan}\ \emph {et~al.}(2015)\citenamefont {Fan},
  \citenamefont {Grychtol}, \citenamefont {Knut}, \citenamefont
  {Hern{\ifmmode\acute{a}\else\'{a}\fi}ndez-Garc{\ifmmode\acute{\imath}\else\'{\i}\fi}a},
  \citenamefont {Hickstein}, \citenamefont {Zusin}, \citenamefont {Gentry},
  \citenamefont {Dollar}, \citenamefont {Mancuso}, \citenamefont {Hogle},
  \citenamefont {Kfir}, \citenamefont {Legut}, \citenamefont {Carva},
  \citenamefont {Ellis}, \citenamefont {Dorney}, \citenamefont {Chen},
  \citenamefont {Shpyrko}, \citenamefont {Fullerton}, \citenamefont {Cohen},
  \citenamefont {Oppeneer}, \citenamefont
  {Milo{\ifmmode\check{s}\else\v{s}\fi}evi{\ifmmode\acute{c}\else\'{c}\fi}},
  \citenamefont {Becker}, \citenamefont
  {Jaro{\ifmmode\acute{n}\else\'{n}\fi}-Becker}, \citenamefont {Popmintchev},
  \citenamefont {Murnane},\ and\ \citenamefont {Kapteyn}}]{Fan2015Nov}%
  \BibitemOpen
  \bibfield  {author} {\bibinfo {author} {\bibfnamefont {T.}~\bibnamefont
  {Fan}}, \bibinfo {author} {\bibfnamefont {P.}~\bibnamefont {Grychtol}},
  \bibinfo {author} {\bibfnamefont {R.}~\bibnamefont {Knut}}, \bibinfo {author}
  {\bibfnamefont {C.}~\bibnamefont
  {Hern{\ifmmode\acute{a}\else\'{a}\fi}ndez-Garc{\ifmmode\acute{\imath}\else\'{\i}\fi}a}},
  \bibinfo {author} {\bibfnamefont {D.~D.}\ \bibnamefont {Hickstein}}, \bibinfo
  {author} {\bibfnamefont {D.}~\bibnamefont {Zusin}}, \bibinfo {author}
  {\bibfnamefont {C.}~\bibnamefont {Gentry}}, \bibinfo {author} {\bibfnamefont
  {F.~J.}\ \bibnamefont {Dollar}}, \bibinfo {author} {\bibfnamefont {C.~A.}\
  \bibnamefont {Mancuso}}, \bibinfo {author} {\bibfnamefont {C.~W.}\
  \bibnamefont {Hogle}}, \bibinfo {author} {\bibfnamefont {O.}~\bibnamefont
  {Kfir}}, \bibinfo {author} {\bibfnamefont {D.}~\bibnamefont {Legut}},
  \bibinfo {author} {\bibfnamefont {K.}~\bibnamefont {Carva}}, \bibinfo
  {author} {\bibfnamefont {J.~L.}\ \bibnamefont {Ellis}}, \bibinfo {author}
  {\bibfnamefont {K.~M.}\ \bibnamefont {Dorney}}, \bibinfo {author}
  {\bibfnamefont {C.}~\bibnamefont {Chen}}, \bibinfo {author} {\bibfnamefont
  {O.~G.}\ \bibnamefont {Shpyrko}}, \bibinfo {author} {\bibfnamefont {E.~E.}\
  \bibnamefont {Fullerton}}, \bibinfo {author} {\bibfnamefont {O.}~\bibnamefont
  {Cohen}}, \bibinfo {author} {\bibfnamefont {P.~M.}\ \bibnamefont {Oppeneer}},
  \bibinfo {author} {\bibfnamefont {D.~B.}\ \bibnamefont
  {Milo{\ifmmode\check{s}\else\v{s}\fi}evi{\ifmmode\acute{c}\else\'{c}\fi}}},
  \bibinfo {author} {\bibfnamefont {A.}~\bibnamefont {Becker}}, \bibinfo
  {author} {\bibfnamefont {A.~A.}\ \bibnamefont
  {Jaro{\ifmmode\acute{n}\else\'{n}\fi}-Becker}}, \bibinfo {author}
  {\bibfnamefont {T.}~\bibnamefont {Popmintchev}}, \bibinfo {author}
  {\bibfnamefont {M.~M.}\ \bibnamefont {Murnane}},\ and\ \bibinfo {author}
  {\bibfnamefont {H.~C.}\ \bibnamefont {Kapteyn}},\ }\bibfield  {title}
  {\bibinfo {title} {{Bright circularly polarized soft X-ray high harmonics for
  X-ray magnetic circular dichroism}},\ }\href
  {https://doi.org/10.1073/pnas.1519666112} {\bibfield  {journal} {\bibinfo
  {journal} {Proc. Natl. Acad. Sci. U.S.A.}\ }\textbf {\bibinfo {volume}
  {112}},\ \bibinfo {pages} {14206} (\bibinfo {year} {2015})}\BibitemShut
  {NoStop}%
\bibitem [{\citenamefont {Lambert}\ \emph {et~al.}(2015)\citenamefont
  {Lambert}, \citenamefont {Vodungbo}, \citenamefont {Gautier}, \citenamefont
  {Mahieu}, \citenamefont {Malka}, \citenamefont {Sebban}, \citenamefont
  {Zeitoun}, \citenamefont {Luning}, \citenamefont {Perron}, \citenamefont
  {Andreev}, \citenamefont {Stremoukhov}, \citenamefont {Ardana-Lamas},
  \citenamefont {Dax}, \citenamefont {Hauri}, \citenamefont {Sardinha},\ and\
  \citenamefont {Fajardo}}]{Lambert_bicircular}%
  \BibitemOpen
  \bibfield  {author} {\bibinfo {author} {\bibfnamefont {G.}~\bibnamefont
  {Lambert}}, \bibinfo {author} {\bibfnamefont {B.}~\bibnamefont {Vodungbo}},
  \bibinfo {author} {\bibfnamefont {J.}~\bibnamefont {Gautier}}, \bibinfo
  {author} {\bibfnamefont {B.}~\bibnamefont {Mahieu}}, \bibinfo {author}
  {\bibfnamefont {V.}~\bibnamefont {Malka}}, \bibinfo {author} {\bibfnamefont
  {S.}~\bibnamefont {Sebban}}, \bibinfo {author} {\bibfnamefont
  {P.}~\bibnamefont {Zeitoun}}, \bibinfo {author} {\bibfnamefont
  {J.}~\bibnamefont {Luning}}, \bibinfo {author} {\bibfnamefont
  {J.}~\bibnamefont {Perron}}, \bibinfo {author} {\bibfnamefont
  {A.}~\bibnamefont {Andreev}}, \bibinfo {author} {\bibfnamefont
  {S.}~\bibnamefont {Stremoukhov}}, \bibinfo {author} {\bibfnamefont
  {F.}~\bibnamefont {Ardana-Lamas}}, \bibinfo {author} {\bibfnamefont
  {A.}~\bibnamefont {Dax}}, \bibinfo {author} {\bibfnamefont {C.~P.}\
  \bibnamefont {Hauri}}, \bibinfo {author} {\bibfnamefont {A.}~\bibnamefont
  {Sardinha}},\ and\ \bibinfo {author} {\bibfnamefont {M.}~\bibnamefont
  {Fajardo}},\ }\bibfield  {title} {\bibinfo {title} {{Towards enabling
  femtosecond helicity-dependent spectroscopy with high-harmonic sources}},\
  }\href {https://doi.org/10.1038/ncomms7167} {\bibfield  {journal} {\bibinfo
  {journal} {Nat. Commun.}\ }\textbf {\bibinfo {volume} {6}},\ \bibinfo {pages}
  {6167} (\bibinfo {year} {2015})}\BibitemShut {NoStop}%
\bibitem [{\citenamefont {Hern\'andez-Garc\'{\i}a}\ \emph
  {et~al.}(2016)\citenamefont {Hern\'andez-Garc\'{\i}a}, \citenamefont
  {Durfee}, \citenamefont {Hickstein}, \citenamefont {Popmintchev},
  \citenamefont {Meier}, \citenamefont {Murnane}, \citenamefont {Kapteyn},
  \citenamefont {Sola}, \citenamefont {Jaron-Becker},\ and\ \citenamefont
  {Becker}}]{Murnane_PRA}%
  \BibitemOpen
  \bibfield  {author} {\bibinfo {author} {\bibfnamefont {C.}~\bibnamefont
  {Hern\'andez-Garc\'{\i}a}}, \bibinfo {author} {\bibfnamefont {C.~G.}\
  \bibnamefont {Durfee}}, \bibinfo {author} {\bibfnamefont {D.~D.}\
  \bibnamefont {Hickstein}}, \bibinfo {author} {\bibfnamefont {T.}~\bibnamefont
  {Popmintchev}}, \bibinfo {author} {\bibfnamefont {A.}~\bibnamefont {Meier}},
  \bibinfo {author} {\bibfnamefont {M.~M.}\ \bibnamefont {Murnane}}, \bibinfo
  {author} {\bibfnamefont {H.~C.}\ \bibnamefont {Kapteyn}}, \bibinfo {author}
  {\bibfnamefont {I.~J.}\ \bibnamefont {Sola}}, \bibinfo {author}
  {\bibfnamefont {A.}~\bibnamefont {Jaron-Becker}},\ and\ \bibinfo {author}
  {\bibfnamefont {A.}~\bibnamefont {Becker}},\ }\bibfield  {title} {\bibinfo
  {title} {Schemes for generation of isolated attosecond pulses of pure
  circular polarization},\ }\href {https://doi.org/10.1103/PhysRevA.93.043855}
  {\bibfield  {journal} {\bibinfo  {journal} {Phys. Rev. A}\ }\textbf {\bibinfo
  {volume} {93}},\ \bibinfo {pages} {043855} (\bibinfo {year}
  {2016})}\BibitemShut {NoStop}%
\bibitem [{\citenamefont {Yuan}\ and\ \citenamefont
  {Bandrauk}(2013)}]{Bandrauk_bicircular}%
  \BibitemOpen
  \bibfield  {author} {\bibinfo {author} {\bibfnamefont {K.-J.}\ \bibnamefont
  {Yuan}}\ and\ \bibinfo {author} {\bibfnamefont {A.~D.}\ \bibnamefont
  {Bandrauk}},\ }\bibfield  {title} {\bibinfo {title} {Single circularly
  polarized attosecond pulse generation by intense few cycle elliptically
  polarized laser pulses and terahertz fields from molecular media},\ }\href
  {https://doi.org/10.1103/PhysRevLett.110.023003} {\bibfield  {journal}
  {\bibinfo  {journal} {Phys. Rev. Lett.}\ }\textbf {\bibinfo {volume} {110}},\
  \bibinfo {pages} {023003} (\bibinfo {year} {2013})}\BibitemShut {NoStop}%
\bibitem [{\citenamefont {Vodungbo}\ \emph {et~al.}(2011)\citenamefont
  {Vodungbo}, \citenamefont {Sardinha}, \citenamefont {Gautier}, \citenamefont
  {Lambert}, \citenamefont {Valentin}, \citenamefont {Lozano}, \citenamefont
  {Iaquaniello}, \citenamefont {Delmotte}, \citenamefont {Sebban},
  \citenamefont {L{\ifmmode\ddot{u}\else\"{u}\fi}ning},\ and\ \citenamefont
  {Zeitoun}}]{Vodungbo2011Feb}%
  \BibitemOpen
  \bibfield  {author} {\bibinfo {author} {\bibfnamefont {B.}~\bibnamefont
  {Vodungbo}}, \bibinfo {author} {\bibfnamefont {A.~B.}\ \bibnamefont
  {Sardinha}}, \bibinfo {author} {\bibfnamefont {J.}~\bibnamefont {Gautier}},
  \bibinfo {author} {\bibfnamefont {G.}~\bibnamefont {Lambert}}, \bibinfo
  {author} {\bibfnamefont {C.}~\bibnamefont {Valentin}}, \bibinfo {author}
  {\bibfnamefont {M.}~\bibnamefont {Lozano}}, \bibinfo {author} {\bibfnamefont
  {G.}~\bibnamefont {Iaquaniello}}, \bibinfo {author} {\bibfnamefont
  {F.}~\bibnamefont {Delmotte}}, \bibinfo {author} {\bibfnamefont
  {S.}~\bibnamefont {Sebban}}, \bibinfo {author} {\bibfnamefont
  {J.}~\bibnamefont {L{\ifmmode\ddot{u}\else\"{u}\fi}ning}},\ and\ \bibinfo
  {author} {\bibfnamefont {P.}~\bibnamefont {Zeitoun}},\ }\bibfield  {title}
  {\bibinfo {title} {{Polarization control of high order harmonics in the EUV
  photon energy range}},\ }\href {https://doi.org/10.1364/OE.19.004346}
  {\bibfield  {journal} {\bibinfo  {journal} {Opt. Express}\ }\textbf {\bibinfo
  {volume} {19}},\ \bibinfo {pages} {4346} (\bibinfo {year}
  {2011})}\BibitemShut {NoStop}%
\bibitem [{\citenamefont {Depresseux}\ \emph {et~al.}(2015)\citenamefont
  {Depresseux}, \citenamefont {Oliva}, \citenamefont {Gautier}, \citenamefont
  {Tissandier}, \citenamefont {Lambert}, \citenamefont {Vodungbo},
  \citenamefont {Goddet}, \citenamefont {Tafzi}, \citenamefont {Nejdl},
  \citenamefont {Kozlova}, \citenamefont {Maynard}, \citenamefont {Kim},
  \citenamefont {Ta~Phuoc}, \citenamefont {Rousse}, \citenamefont {Zeitoun},\
  and\ \citenamefont {Sebban}}]{PhysRevLett.115.083901}%
  \BibitemOpen
  \bibfield  {author} {\bibinfo {author} {\bibfnamefont {A.}~\bibnamefont
  {Depresseux}}, \bibinfo {author} {\bibfnamefont {E.}~\bibnamefont {Oliva}},
  \bibinfo {author} {\bibfnamefont {J.}~\bibnamefont {Gautier}}, \bibinfo
  {author} {\bibfnamefont {F.}~\bibnamefont {Tissandier}}, \bibinfo {author}
  {\bibfnamefont {G.}~\bibnamefont {Lambert}}, \bibinfo {author} {\bibfnamefont
  {B.}~\bibnamefont {Vodungbo}}, \bibinfo {author} {\bibfnamefont {J.-P.}\
  \bibnamefont {Goddet}}, \bibinfo {author} {\bibfnamefont {A.}~\bibnamefont
  {Tafzi}}, \bibinfo {author} {\bibfnamefont {J.}~\bibnamefont {Nejdl}},
  \bibinfo {author} {\bibfnamefont {M.}~\bibnamefont {Kozlova}}, \bibinfo
  {author} {\bibfnamefont {G.}~\bibnamefont {Maynard}}, \bibinfo {author}
  {\bibfnamefont {H.~T.}\ \bibnamefont {Kim}}, \bibinfo {author} {\bibfnamefont
  {K.}~\bibnamefont {Ta~Phuoc}}, \bibinfo {author} {\bibfnamefont
  {A.}~\bibnamefont {Rousse}}, \bibinfo {author} {\bibfnamefont
  {P.}~\bibnamefont {Zeitoun}},\ and\ \bibinfo {author} {\bibfnamefont
  {S.}~\bibnamefont {Sebban}},\ }\bibfield  {title} {\bibinfo {title}
  {Demonstration of a circularly polarized plasma-based soft-x-ray laser},\
  }\href {https://doi.org/10.1103/PhysRevLett.115.083901} {\bibfield  {journal}
  {\bibinfo  {journal} {Phys. Rev. Lett.}\ }\textbf {\bibinfo {volume} {115}},\
  \bibinfo {pages} {083901} (\bibinfo {year} {2015})}\BibitemShut {NoStop}%
\bibitem [{\citenamefont {Schmidt}\ \emph {et~al.}(2015)\citenamefont
  {Schmidt}, \citenamefont {Guggenmos}, \citenamefont {Hofstetter},
  \citenamefont {Chew},\ and\ \citenamefont {Kleineberg}}]{Schmidt2015Dec}%
  \BibitemOpen
  \bibfield  {author} {\bibinfo {author} {\bibfnamefont {J.}~\bibnamefont
  {Schmidt}}, \bibinfo {author} {\bibfnamefont {A.}~\bibnamefont {Guggenmos}},
  \bibinfo {author} {\bibfnamefont {M.}~\bibnamefont {Hofstetter}}, \bibinfo
  {author} {\bibfnamefont {S.~H.}\ \bibnamefont {Chew}},\ and\ \bibinfo
  {author} {\bibfnamefont {U.}~\bibnamefont {Kleineberg}},\ }\bibfield  {title}
  {\bibinfo {title} {{Generation of circularly polarized high harmonic
  radiation using a transmission multilayer quarter waveplate}},\ }\href
  {https://doi.org/10.1364/OE.23.033564} {\bibfield  {journal} {\bibinfo
  {journal} {Opt. Express}\ }\textbf {\bibinfo {volume} {23}},\ \bibinfo
  {pages} {33564} (\bibinfo {year} {2015})}\BibitemShut {NoStop}%
\bibitem [{\citenamefont {Azoury}\ \emph {et~al.}(2019)\citenamefont {Azoury},
  \citenamefont {Kneller}, \citenamefont
  {Kr{\ifmmode\ddot{u}\else\"{u}\fi}ger}, \citenamefont {Bruner}, \citenamefont
  {Cohen}, \citenamefont {Mairesse},\ and\ \citenamefont
  {Dudovich}}]{Azoury2019Feb}%
  \BibitemOpen
  \bibfield  {author} {\bibinfo {author} {\bibfnamefont {D.}~\bibnamefont
  {Azoury}}, \bibinfo {author} {\bibfnamefont {O.}~\bibnamefont {Kneller}},
  \bibinfo {author} {\bibfnamefont {M.}~\bibnamefont
  {Kr{\ifmmode\ddot{u}\else\"{u}\fi}ger}}, \bibinfo {author} {\bibfnamefont
  {B.~D.}\ \bibnamefont {Bruner}}, \bibinfo {author} {\bibfnamefont
  {O.}~\bibnamefont {Cohen}}, \bibinfo {author} {\bibfnamefont
  {Y.}~\bibnamefont {Mairesse}},\ and\ \bibinfo {author} {\bibfnamefont
  {N.}~\bibnamefont {Dudovich}},\ }\bibfield  {title} {\bibinfo {title}
  {{Interferometric attosecond lock-in measurement of extreme-ultraviolet
  circular dichroism}},\ }\href {https://doi.org/10.1038/s41566-019-0350-5}
  {\bibfield  {journal} {\bibinfo  {journal} {Nat. Photonics}\ }\textbf
  {\bibinfo {volume} {13}},\ \bibinfo {pages} {198} (\bibinfo {year}
  {2019})}\BibitemShut {NoStop}%
\bibitem [{\citenamefont {Ellis}\ \emph {et~al.}(2018)\citenamefont {Ellis},
  \citenamefont {Dorney}, \citenamefont {Hickstein}, \citenamefont {Brooks},
  \citenamefont {Gentry}, \citenamefont
  {Hern{\ifmmode\acute{a}\else\'{a}\fi}ndez-Garc{\ifmmode\acute{\imath}\else\'{\i}\fi}a},
  \citenamefont {Zusin}, \citenamefont {Shaw}, \citenamefont {Nguyen},
  \citenamefont {Mancuso}, \citenamefont {Jansen}, \citenamefont {Witte},
  \citenamefont {Kapteyn},\ and\ \citenamefont {Murnane}}]{Ellis2018}%
  \BibitemOpen
  \bibfield  {author} {\bibinfo {author} {\bibfnamefont {J.~L.}\ \bibnamefont
  {Ellis}}, \bibinfo {author} {\bibfnamefont {K.~M.}\ \bibnamefont {Dorney}},
  \bibinfo {author} {\bibfnamefont {D.~D.}\ \bibnamefont {Hickstein}}, \bibinfo
  {author} {\bibfnamefont {N.~J.}\ \bibnamefont {Brooks}}, \bibinfo {author}
  {\bibfnamefont {C.}~\bibnamefont {Gentry}}, \bibinfo {author} {\bibfnamefont
  {C.}~\bibnamefont
  {Hern{\ifmmode\acute{a}\else\'{a}\fi}ndez-Garc{\ifmmode\acute{\imath}\else\'{\i}\fi}a}},
  \bibinfo {author} {\bibfnamefont {D.}~\bibnamefont {Zusin}}, \bibinfo
  {author} {\bibfnamefont {J.~M.}\ \bibnamefont {Shaw}}, \bibinfo {author}
  {\bibfnamefont {Q.~L.}\ \bibnamefont {Nguyen}}, \bibinfo {author}
  {\bibfnamefont {C.~A.}\ \bibnamefont {Mancuso}}, \bibinfo {author}
  {\bibfnamefont {G.~S.~M.}\ \bibnamefont {Jansen}}, \bibinfo {author}
  {\bibfnamefont {S.}~\bibnamefont {Witte}}, \bibinfo {author} {\bibfnamefont
  {H.~C.}\ \bibnamefont {Kapteyn}},\ and\ \bibinfo {author} {\bibfnamefont
  {M.~M.}\ \bibnamefont {Murnane}},\ }\bibfield  {title} {\bibinfo {title}
  {{High harmonics with spatially varying ellipticity}},\ }\href
  {https://doi.org/10.1364/OPTICA.5.000479} {\bibfield  {journal} {\bibinfo
  {journal} {Optica}\ }\textbf {\bibinfo {volume} {5}},\ \bibinfo {pages} {479}
  (\bibinfo {year} {2018})}\BibitemShut {NoStop}%
\bibitem [{\citenamefont {Huang}\ \emph {et~al.}(2018)\citenamefont {Huang},
  \citenamefont
  {Hern{\ifmmode\acute{a}\else\'{a}\fi}ndez-Garc{\ifmmode\acute{\imath}\else\'{\i}\fi}a},
  \citenamefont {Huang}, \citenamefont {Huang}, \citenamefont {Lu},
  \citenamefont {Rego}, \citenamefont {Hickstein}, \citenamefont {Ellis},
  \citenamefont {Jaron-Becker}, \citenamefont {Becker}, \citenamefont {Yang},
  \citenamefont {Durfee}, \citenamefont {Plaja}, \citenamefont {Kapteyn},
  \citenamefont {Murnane}, \citenamefont {Kung},\ and\ \citenamefont
  {Chen}}]{Huang2018}%
  \BibitemOpen
  \bibfield  {author} {\bibinfo {author} {\bibfnamefont {P.-C.}\ \bibnamefont
  {Huang}}, \bibinfo {author} {\bibfnamefont {C.}~\bibnamefont
  {Hern{\ifmmode\acute{a}\else\'{a}\fi}ndez-Garc{\ifmmode\acute{\imath}\else\'{\i}\fi}a}},
  \bibinfo {author} {\bibfnamefont {J.-T.}\ \bibnamefont {Huang}}, \bibinfo
  {author} {\bibfnamefont {P.-Y.}\ \bibnamefont {Huang}}, \bibinfo {author}
  {\bibfnamefont {C.-H.}\ \bibnamefont {Lu}}, \bibinfo {author} {\bibfnamefont
  {L.}~\bibnamefont {Rego}}, \bibinfo {author} {\bibfnamefont {D.~D.}\
  \bibnamefont {Hickstein}}, \bibinfo {author} {\bibfnamefont {J.~L.}\
  \bibnamefont {Ellis}}, \bibinfo {author} {\bibfnamefont {A.}~\bibnamefont
  {Jaron-Becker}}, \bibinfo {author} {\bibfnamefont {A.}~\bibnamefont
  {Becker}}, \bibinfo {author} {\bibfnamefont {S.-D.}\ \bibnamefont {Yang}},
  \bibinfo {author} {\bibfnamefont {C.~G.}\ \bibnamefont {Durfee}}, \bibinfo
  {author} {\bibfnamefont {L.}~\bibnamefont {Plaja}}, \bibinfo {author}
  {\bibfnamefont {H.~C.}\ \bibnamefont {Kapteyn}}, \bibinfo {author}
  {\bibfnamefont {M.~M.}\ \bibnamefont {Murnane}}, \bibinfo {author}
  {\bibfnamefont {A.~H.}\ \bibnamefont {Kung}},\ and\ \bibinfo {author}
  {\bibfnamefont {M.-C.}\ \bibnamefont {Chen}},\ }\bibfield  {title} {\bibinfo
  {title} {{Polarization control of isolated high-harmonic pulses}},\ }\href
  {https://doi.org/10.1038/s41566-018-0145-0} {\bibfield  {journal} {\bibinfo
  {journal} {Nat. Photonics}\ }\textbf {\bibinfo {volume} {12}},\ \bibinfo
  {pages} {349} (\bibinfo {year} {2018})}\BibitemShut {NoStop}%
\bibitem [{\citenamefont {Zhou}\ \emph {et~al.}(2009)\citenamefont {Zhou},
  \citenamefont {Lock}, \citenamefont {Wagner}, \citenamefont {Li},
  \citenamefont {Kapteyn},\ and\ \citenamefont
  {Murnane}}]{PhysRevLett.102.073902}%
  \BibitemOpen
  \bibfield  {author} {\bibinfo {author} {\bibfnamefont {X.}~\bibnamefont
  {Zhou}}, \bibinfo {author} {\bibfnamefont {R.}~\bibnamefont {Lock}}, \bibinfo
  {author} {\bibfnamefont {N.}~\bibnamefont {Wagner}}, \bibinfo {author}
  {\bibfnamefont {W.}~\bibnamefont {Li}}, \bibinfo {author} {\bibfnamefont
  {H.~C.}\ \bibnamefont {Kapteyn}},\ and\ \bibinfo {author} {\bibfnamefont
  {M.~M.}\ \bibnamefont {Murnane}},\ }\bibfield  {title} {\bibinfo {title}
  {Elliptically polarized high-order harmonic emission from molecules in
  linearly polarized laser fields},\ }\href
  {https://doi.org/10.1103/PhysRevLett.102.073902} {\bibfield  {journal}
  {\bibinfo  {journal} {Phys. Rev. Lett.}\ }\textbf {\bibinfo {volume} {102}},\
  \bibinfo {pages} {073902} (\bibinfo {year} {2009})}\BibitemShut {NoStop}%
\bibitem [{\citenamefont {Mairesse}\ \emph {et~al.}(2010)\citenamefont
  {Mairesse}, \citenamefont {Higuet}, \citenamefont {Dudovich}, \citenamefont
  {Shafir}, \citenamefont {Fabre}, \citenamefont {M\'evel}, \citenamefont
  {Constant}, \citenamefont {Patchkovskii}, \citenamefont {Walters},
  \citenamefont {Ivanov},\ and\ \citenamefont
  {Smirnova}}]{PhysRevLett.104.213601}%
  \BibitemOpen
  \bibfield  {author} {\bibinfo {author} {\bibfnamefont {Y.}~\bibnamefont
  {Mairesse}}, \bibinfo {author} {\bibfnamefont {J.}~\bibnamefont {Higuet}},
  \bibinfo {author} {\bibfnamefont {N.}~\bibnamefont {Dudovich}}, \bibinfo
  {author} {\bibfnamefont {D.}~\bibnamefont {Shafir}}, \bibinfo {author}
  {\bibfnamefont {B.}~\bibnamefont {Fabre}}, \bibinfo {author} {\bibfnamefont
  {E.}~\bibnamefont {M\'evel}}, \bibinfo {author} {\bibfnamefont
  {E.}~\bibnamefont {Constant}}, \bibinfo {author} {\bibfnamefont
  {S.}~\bibnamefont {Patchkovskii}}, \bibinfo {author} {\bibfnamefont
  {Z.}~\bibnamefont {Walters}}, \bibinfo {author} {\bibfnamefont {M.~Y.}\
  \bibnamefont {Ivanov}},\ and\ \bibinfo {author} {\bibfnamefont
  {O.}~\bibnamefont {Smirnova}},\ }\bibfield  {title} {\bibinfo {title} {High
  harmonic spectroscopy of multichannel dynamics in strong-field ionization},\
  }\href {https://doi.org/10.1103/PhysRevLett.104.213601} {\bibfield  {journal}
  {\bibinfo  {journal} {Phys. Rev. Lett.}\ }\textbf {\bibinfo {volume} {104}},\
  \bibinfo {pages} {213601} (\bibinfo {year} {2010})}\BibitemShut {NoStop}%
\bibitem [{\citenamefont {Skantzakis}\ \emph {et~al.}(2016)\citenamefont
  {Skantzakis}, \citenamefont {Chatziathanasiou}, \citenamefont {Carpeggiani},
  \citenamefont {Sansone}, \citenamefont {Nayak}, \citenamefont {Gray},
  \citenamefont {Tzallas}, \citenamefont {Charalambidis}, \citenamefont
  {Hertz},\ and\ \citenamefont {Faucher}}]{Skantzakis2016Dec}%
  \BibitemOpen
  \bibfield  {author} {\bibinfo {author} {\bibfnamefont {E.}~\bibnamefont
  {Skantzakis}}, \bibinfo {author} {\bibfnamefont {S.}~\bibnamefont
  {Chatziathanasiou}}, \bibinfo {author} {\bibfnamefont {P.~A.}\ \bibnamefont
  {Carpeggiani}}, \bibinfo {author} {\bibfnamefont {G.}~\bibnamefont
  {Sansone}}, \bibinfo {author} {\bibfnamefont {A.}~\bibnamefont {Nayak}},
  \bibinfo {author} {\bibfnamefont {D.}~\bibnamefont {Gray}}, \bibinfo {author}
  {\bibfnamefont {P.}~\bibnamefont {Tzallas}}, \bibinfo {author} {\bibfnamefont
  {D.}~\bibnamefont {Charalambidis}}, \bibinfo {author} {\bibfnamefont
  {E.}~\bibnamefont {Hertz}},\ and\ \bibinfo {author} {\bibfnamefont
  {O.}~\bibnamefont {Faucher}},\ }\bibfield  {title} {\bibinfo {title}
  {{Polarization shaping of high-order harmonics in laser-aligned molecules}},\
  }\href {https://doi.org/10.1038/srep39295} {\bibfield  {journal} {\bibinfo
  {journal} {Sci. Rep.}\ }\textbf {\bibinfo {volume} {6}},\ \bibinfo {pages}
  {1} (\bibinfo {year} {2016})}\BibitemShut {NoStop}%
\bibitem [{\citenamefont {Ganeev}\ \emph {et~al.}(2006)\citenamefont {Ganeev},
  \citenamefont {Suzuki}, \citenamefont {Baba}, \citenamefont {Kuroda},\ and\
  \citenamefont {Ozaki}}]{Ganeev2006Jun}%
  \BibitemOpen
  \bibfield  {author} {\bibinfo {author} {\bibfnamefont {R.~A.}\ \bibnamefont
  {Ganeev}}, \bibinfo {author} {\bibfnamefont {M.}~\bibnamefont {Suzuki}},
  \bibinfo {author} {\bibfnamefont {M.}~\bibnamefont {Baba}}, \bibinfo {author}
  {\bibfnamefont {H.}~\bibnamefont {Kuroda}},\ and\ \bibinfo {author}
  {\bibfnamefont {T.}~\bibnamefont {Ozaki}},\ }\bibfield  {title} {\bibinfo
  {title} {{Strong resonance enhancement of a single harmonic generated in the
  extreme ultraviolet range}},\ }\href {https://doi.org/10.1364/OL.31.001699}
  {\bibfield  {journal} {\bibinfo  {journal} {Opt. Lett.}\ }\textbf {\bibinfo
  {volume} {31}},\ \bibinfo {pages} {1699} (\bibinfo {year}
  {2006})}\BibitemShut {NoStop}%
\bibitem [{\citenamefont {Frolov}\ \emph {et~al.}(2009)\citenamefont {Frolov},
  \citenamefont {Manakov}, \citenamefont {Sarantseva}, \citenamefont {Emelin},
  \citenamefont {Ryabikin},\ and\ \citenamefont
  {Starace}}]{PhysRevLett.102.243901}%
  \BibitemOpen
  \bibfield  {author} {\bibinfo {author} {\bibfnamefont {M.~V.}\ \bibnamefont
  {Frolov}}, \bibinfo {author} {\bibfnamefont {N.~L.}\ \bibnamefont {Manakov}},
  \bibinfo {author} {\bibfnamefont {T.~S.}\ \bibnamefont {Sarantseva}},
  \bibinfo {author} {\bibfnamefont {M.~Y.}\ \bibnamefont {Emelin}}, \bibinfo
  {author} {\bibfnamefont {M.~Y.}\ \bibnamefont {Ryabikin}},\ and\ \bibinfo
  {author} {\bibfnamefont {A.~F.}\ \bibnamefont {Starace}},\ }\bibfield
  {title} {\bibinfo {title} {Analytic description of the high-energy plateau in
  harmonic generation by atoms: Can the harmonic power increase with increasing
  laser wavelengths?},\ }\href {https://doi.org/10.1103/PhysRevLett.102.243901}
  {\bibfield  {journal} {\bibinfo  {journal} {Phys. Rev. Lett.}\ }\textbf
  {\bibinfo {volume} {102}},\ \bibinfo {pages} {243901} (\bibinfo {year}
  {2009})}\BibitemShut {NoStop}%
\bibitem [{\citenamefont {Strelkov}(2010)}]{Strelkov_PRL_104}%
  \BibitemOpen
  \bibfield  {author} {\bibinfo {author} {\bibfnamefont {V.}~\bibnamefont
  {Strelkov}},\ }\bibfield  {title} {\bibinfo {title} {Role of autoionizing
  state in resonant high-order harmonic generation and attosecond pulse
  production},\ }\href {https://doi.org/10.1103/PhysRevLett.104.123901}
  {\bibfield  {journal} {\bibinfo  {journal} {Phys. Rev. Lett.}\ }\textbf
  {\bibinfo {volume} {104}},\ \bibinfo {pages} {123901} (\bibinfo {year}
  {2010})}\BibitemShut {NoStop}%
\bibitem [{\citenamefont {Shiner}\ \emph {et~al.}(2011)\citenamefont {Shiner},
  \citenamefont {Schmidt}, \citenamefont {Trallero-Herrero}, \citenamefont
  {W{\ifmmode\ddot{o}\else\"{o}\fi}rner}, \citenamefont {Patchkovskii},
  \citenamefont {Corkum}, \citenamefont {Kieffer}, \citenamefont
  {L{\ifmmode\acute{e}\else\'{e}\fi}gar{\ifmmode\acute{e}\else\'{e}\fi}},\ and\
  \citenamefont {Villeneuve}}]{Shiner2011}%
  \BibitemOpen
  \bibfield  {author} {\bibinfo {author} {\bibfnamefont {A.~D.}\ \bibnamefont
  {Shiner}}, \bibinfo {author} {\bibfnamefont {B.~E.}\ \bibnamefont {Schmidt}},
  \bibinfo {author} {\bibfnamefont {C.}~\bibnamefont {Trallero-Herrero}},
  \bibinfo {author} {\bibfnamefont {H.~J.}\ \bibnamefont
  {W{\ifmmode\ddot{o}\else\"{o}\fi}rner}}, \bibinfo {author} {\bibfnamefont
  {S.}~\bibnamefont {Patchkovskii}}, \bibinfo {author} {\bibfnamefont {P.~B.}\
  \bibnamefont {Corkum}}, \bibinfo {author} {\bibfnamefont {J.-C.}\
  \bibnamefont {Kieffer}}, \bibinfo {author} {\bibfnamefont {F.}~\bibnamefont
  {L{\ifmmode\acute{e}\else\'{e}\fi}gar{\ifmmode\acute{e}\else\'{e}\fi}}},\
  and\ \bibinfo {author} {\bibfnamefont {D.~M.}\ \bibnamefont {Villeneuve}},\
  }\bibfield  {title} {\bibinfo {title} {{Probing collective multi-electron
  dynamics in xenon with high-harmonic spectroscopy}},\ }\href
  {https://doi.org/10.1038/nphys1940} {\bibfield  {journal} {\bibinfo
  {journal} {Nat. Phys.}\ }\textbf {\bibinfo {volume} {7}},\ \bibinfo {pages}
  {464} (\bibinfo {year} {2011})}\BibitemShut {NoStop}%
\bibitem [{\citenamefont {Strelkov}\ \emph {et~al.}(2014)\citenamefont
  {Strelkov}, \citenamefont {Khokhlova},\ and\ \citenamefont
  {Shubin}}]{Strelkov_Fano}%
  \BibitemOpen
  \bibfield  {author} {\bibinfo {author} {\bibfnamefont {V.~V.}\ \bibnamefont
  {Strelkov}}, \bibinfo {author} {\bibfnamefont {M.~A.}\ \bibnamefont
  {Khokhlova}},\ and\ \bibinfo {author} {\bibfnamefont {N.~Y.}\ \bibnamefont
  {Shubin}},\ }\bibfield  {title} {\bibinfo {title} {High-order harmonic
  generation and {Fano} resonances},\ }\href
  {https://doi.org/10.1103/PhysRevA.89.053833} {\bibfield  {journal} {\bibinfo
  {journal} {Phys. Rev. A}\ }\textbf {\bibinfo {volume} {89}},\ \bibinfo
  {pages} {053833} (\bibinfo {year} {2014})}\BibitemShut {NoStop}%
\bibitem [{\citenamefont {Ferr{\ifmmode\acute{e}\else\'{e}\fi}}\ \emph
  {et~al.}(2014)\citenamefont {Ferr{\ifmmode\acute{e}\else\'{e}\fi}},
  \citenamefont {Handschin}, \citenamefont {Dumergue}, \citenamefont {Burgy},
  \citenamefont {Comby}, \citenamefont {Descamps}, \citenamefont {Fabre},
  \citenamefont {Garcia}, \citenamefont
  {G{\ifmmode\acute{e}\else\'{e}\fi}neaux}, \citenamefont {Merceron},
  \citenamefont {M{\ifmmode\acute{e}\else\'{e}\fi}vel}, \citenamefont {Nahon},
  \citenamefont {Petit}, \citenamefont {Pons}, \citenamefont {Staedter},
  \citenamefont {Weber}, \citenamefont {Ruchon}, \citenamefont {Blanchet},\
  and\ \citenamefont {Mairesse}}]{Ferre2014_rhhg}%
  \BibitemOpen
  \bibfield  {author} {\bibinfo {author} {\bibfnamefont {A.}~\bibnamefont
  {Ferr{\ifmmode\acute{e}\else\'{e}\fi}}}, \bibinfo {author} {\bibfnamefont
  {C.}~\bibnamefont {Handschin}}, \bibinfo {author} {\bibfnamefont
  {M.}~\bibnamefont {Dumergue}}, \bibinfo {author} {\bibfnamefont
  {F.}~\bibnamefont {Burgy}}, \bibinfo {author} {\bibfnamefont
  {A.}~\bibnamefont {Comby}}, \bibinfo {author} {\bibfnamefont
  {D.}~\bibnamefont {Descamps}}, \bibinfo {author} {\bibfnamefont
  {B.}~\bibnamefont {Fabre}}, \bibinfo {author} {\bibfnamefont {G.~A.}\
  \bibnamefont {Garcia}}, \bibinfo {author} {\bibfnamefont {R.}~\bibnamefont
  {G{\ifmmode\acute{e}\else\'{e}\fi}neaux}}, \bibinfo {author} {\bibfnamefont
  {L.}~\bibnamefont {Merceron}}, \bibinfo {author} {\bibfnamefont
  {E.}~\bibnamefont {M{\ifmmode\acute{e}\else\'{e}\fi}vel}}, \bibinfo {author}
  {\bibfnamefont {L.}~\bibnamefont {Nahon}}, \bibinfo {author} {\bibfnamefont
  {S.}~\bibnamefont {Petit}}, \bibinfo {author} {\bibfnamefont
  {B.}~\bibnamefont {Pons}}, \bibinfo {author} {\bibfnamefont {D.}~\bibnamefont
  {Staedter}}, \bibinfo {author} {\bibfnamefont {S.}~\bibnamefont {Weber}},
  \bibinfo {author} {\bibfnamefont {T.}~\bibnamefont {Ruchon}}, \bibinfo
  {author} {\bibfnamefont {V.}~\bibnamefont {Blanchet}},\ and\ \bibinfo
  {author} {\bibfnamefont {Y.}~\bibnamefont {Mairesse}},\ }\bibfield  {title}
  {\bibinfo {title} {{A table-top ultrashort light source in the extreme
  ultraviolet for circular dichroism experiments}},\ }\href
  {https://doi.org/10.1038/nphoton.2014.314} {\bibfield  {journal} {\bibinfo
  {journal} {Nat. Photonics}\ }\textbf {\bibinfo {volume} {9}},\ \bibinfo
  {pages} {93} (\bibinfo {year} {2014})}\BibitemShut {NoStop}%
\bibitem [{\citenamefont {Beaulieu}\ \emph
  {et~al.}(2016{\natexlab{b}})\citenamefont {Beaulieu}, \citenamefont {Camp},
  \citenamefont {Descamps}, \citenamefont {Comby}, \citenamefont {Wanie},
  \citenamefont {Petit}, \citenamefont
  {L{\ifmmode\acute{e}\else\'{e}\fi}gar{\ifmmode\acute{e}\else\'{e}\fi}},
  \citenamefont {Schafer}, \citenamefont {Gaarde}, \citenamefont {Catoire},\
  and\ \citenamefont {Mairesse}}]{Beaulieu2016}%
  \BibitemOpen
  \bibfield  {author} {\bibinfo {author} {\bibfnamefont {S.}~\bibnamefont
  {Beaulieu}}, \bibinfo {author} {\bibfnamefont {S.}~\bibnamefont {Camp}},
  \bibinfo {author} {\bibfnamefont {D.}~\bibnamefont {Descamps}}, \bibinfo
  {author} {\bibfnamefont {A.}~\bibnamefont {Comby}}, \bibinfo {author}
  {\bibfnamefont {V.}~\bibnamefont {Wanie}}, \bibinfo {author} {\bibfnamefont
  {S.}~\bibnamefont {Petit}}, \bibinfo {author} {\bibfnamefont
  {F.}~\bibnamefont
  {L{\ifmmode\acute{e}\else\'{e}\fi}gar{\ifmmode\acute{e}\else\'{e}\fi}}},
  \bibinfo {author} {\bibfnamefont {K.~J.}\ \bibnamefont {Schafer}}, \bibinfo
  {author} {\bibfnamefont {M.~B.}\ \bibnamefont {Gaarde}}, \bibinfo {author}
  {\bibfnamefont {F.}~\bibnamefont {Catoire}},\ and\ \bibinfo {author}
  {\bibfnamefont {Y.}~\bibnamefont {Mairesse}},\ }\bibfield  {title} {\bibinfo
  {title} {{Role of excited states in high-order harmonic generation}},\ }\href
  {https://doi.org/10.1103/PhysRevLett.117.203001} {\bibfield  {journal}
  {\bibinfo  {journal} {Phys. Rev. Lett.}\ }\textbf {\bibinfo {volume} {117}},\
  \bibinfo {pages} {203001} (\bibinfo {year} {2016}{\natexlab{b}})}\BibitemShut
  {NoStop}%
\bibitem [{\citenamefont {Camp}\ \emph {et~al.}(2018)\citenamefont {Camp},
  \citenamefont {Beaulieu}, \citenamefont {Schafer},\ and\ \citenamefont
  {Gaarde}}]{Camp2018}%
  \BibitemOpen
  \bibfield  {author} {\bibinfo {author} {\bibfnamefont {S.}~\bibnamefont
  {Camp}}, \bibinfo {author} {\bibfnamefont {S.}~\bibnamefont {Beaulieu}},
  \bibinfo {author} {\bibfnamefont {K.~J.}\ \bibnamefont {Schafer}},\ and\
  \bibinfo {author} {\bibfnamefont {M.~B.}\ \bibnamefont {Gaarde}},\ }\bibfield
   {title} {\bibinfo {title} {{Resonantly-initiated quantum trajectories and
  their role in the generation of near-threshold harmonics}},\ }\href
  {https://doi.org/10.1088/1361-6455/aaac12} {\bibfield  {journal} {\bibinfo
  {journal} {J. Phys. B: At. Mol. Opt. Phys.}\ }\textbf {\bibinfo {volume}
  {51}},\ \bibinfo {pages} {064001} (\bibinfo {year} {2018})}\BibitemShut
  {NoStop}%
\bibitem [{\citenamefont {Ferr{\ifmmode\acute{e}\else\'{e}\fi}}\ \emph
  {et~al.}(2015)\citenamefont {Ferr{\ifmmode\acute{e}\else\'{e}\fi}},
  \citenamefont {Boguslavskiy}, \citenamefont {Dagan}, \citenamefont
  {Blanchet}, \citenamefont {Bruner}, \citenamefont {Burgy}, \citenamefont
  {Camper}, \citenamefont {Descamps}, \citenamefont {Fabre}, \citenamefont
  {Fedorov}, \citenamefont {Gaudin}, \citenamefont {Geoffroy}, \citenamefont
  {Mikosch}, \citenamefont {Patchkovskii}, \citenamefont {Petit}, \citenamefont
  {Ruchon}, \citenamefont {Soifer}, \citenamefont {Staedter}, \citenamefont
  {Wilkinson}, \citenamefont {Stolow}, \citenamefont {Dudovich},\ and\
  \citenamefont {Mairesse}}]{Ferre2015Jan}%
  \BibitemOpen
  \bibfield  {author} {\bibinfo {author} {\bibfnamefont {A.}~\bibnamefont
  {Ferr{\ifmmode\acute{e}\else\'{e}\fi}}}, \bibinfo {author} {\bibfnamefont
  {A.~E.}\ \bibnamefont {Boguslavskiy}}, \bibinfo {author} {\bibfnamefont
  {M.}~\bibnamefont {Dagan}}, \bibinfo {author} {\bibfnamefont
  {V.}~\bibnamefont {Blanchet}}, \bibinfo {author} {\bibfnamefont {B.~D.}\
  \bibnamefont {Bruner}}, \bibinfo {author} {\bibfnamefont {F.}~\bibnamefont
  {Burgy}}, \bibinfo {author} {\bibfnamefont {A.}~\bibnamefont {Camper}},
  \bibinfo {author} {\bibfnamefont {D.}~\bibnamefont {Descamps}}, \bibinfo
  {author} {\bibfnamefont {B.}~\bibnamefont {Fabre}}, \bibinfo {author}
  {\bibfnamefont {N.}~\bibnamefont {Fedorov}}, \bibinfo {author} {\bibfnamefont
  {J.}~\bibnamefont {Gaudin}}, \bibinfo {author} {\bibfnamefont
  {G.}~\bibnamefont {Geoffroy}}, \bibinfo {author} {\bibfnamefont
  {J.}~\bibnamefont {Mikosch}}, \bibinfo {author} {\bibfnamefont
  {S.}~\bibnamefont {Patchkovskii}}, \bibinfo {author} {\bibfnamefont
  {S.}~\bibnamefont {Petit}}, \bibinfo {author} {\bibfnamefont
  {T.}~\bibnamefont {Ruchon}}, \bibinfo {author} {\bibfnamefont
  {H.}~\bibnamefont {Soifer}}, \bibinfo {author} {\bibfnamefont
  {D.}~\bibnamefont {Staedter}}, \bibinfo {author} {\bibfnamefont
  {I.}~\bibnamefont {Wilkinson}}, \bibinfo {author} {\bibfnamefont
  {A.}~\bibnamefont {Stolow}}, \bibinfo {author} {\bibfnamefont
  {N.}~\bibnamefont {Dudovich}},\ and\ \bibinfo {author} {\bibfnamefont
  {Y.}~\bibnamefont {Mairesse}},\ }\bibfield  {title} {\bibinfo {title}
  {{Multi-channel electronic and vibrational dynamics in polyatomic resonant
  high-order harmonic generation}},\ }\href
  {https://doi.org/10.1038/ncomms6952} {\bibfield  {journal} {\bibinfo
  {journal} {Nat. Commun.}\ }\textbf {\bibinfo {volume} {6}},\ \bibinfo {pages}
  {1} (\bibinfo {year} {2015})}\BibitemShut {NoStop}%
\bibitem [{\citenamefont {Antoine}\ \emph {et~al.}(1996)\citenamefont
  {Antoine}, \citenamefont {L'Huillier}, \citenamefont {Lewenstein},
  \citenamefont {Sali\`eres},\ and\ \citenamefont {Carr\'e}}]{Antoine_PRA_53}%
  \BibitemOpen
  \bibfield  {author} {\bibinfo {author} {\bibfnamefont {P.}~\bibnamefont
  {Antoine}}, \bibinfo {author} {\bibfnamefont {A.}~\bibnamefont {L'Huillier}},
  \bibinfo {author} {\bibfnamefont {M.}~\bibnamefont {Lewenstein}}, \bibinfo
  {author} {\bibfnamefont {P.}~\bibnamefont {Sali\`eres}},\ and\ \bibinfo
  {author} {\bibfnamefont {B.}~\bibnamefont {Carr\'e}},\ }\bibfield  {title}
  {\bibinfo {title} {Theory of high-order harmonic generation by an
  elliptically polarized laser field},\ }\href
  {https://doi.org/10.1103/PhysRevA.53.1725} {\bibfield  {journal} {\bibinfo
  {journal} {Phys. Rev. A}\ }\textbf {\bibinfo {volume} {53}},\ \bibinfo
  {pages} {1725} (\bibinfo {year} {1996})}\BibitemShut {NoStop}%
\bibitem [{\citenamefont {Fleck}\ \emph {et~al.}(1976)\citenamefont {Fleck},
  \citenamefont {Morris},\ and\ \citenamefont {Feit}}]{Fleck1976}%
  \BibitemOpen
  \bibfield  {author} {\bibinfo {author} {\bibfnamefont {J.~A.}\ \bibnamefont
  {Fleck}}, \bibinfo {author} {\bibfnamefont {J.~R.}\ \bibnamefont {Morris}},\
  and\ \bibinfo {author} {\bibfnamefont {M.~D.}\ \bibnamefont {Feit}},\
  }\bibfield  {title} {\bibinfo {title} {{Time-dependent propagation of high
  energy laser beams through the atmosphere}},\ }\href
  {https://doi.org/10.1007/BF00896333} {\bibfield  {journal} {\bibinfo
  {journal} {Appl. Phys.}\ }\textbf {\bibinfo {volume} {10}},\ \bibinfo {pages}
  {129} (\bibinfo {year} {1976})}\BibitemShut {NoStop}%
\bibitem [{\citenamefont {Ganeev}\ \emph {et~al.}(2012)\citenamefont {Ganeev},
  \citenamefont {Strelkov}, \citenamefont {Hutchison}, \citenamefont
  {Za\"{\i}r}, \citenamefont {Kilbane}, \citenamefont {Khokhlova},\ and\
  \citenamefont {Marangos}}]{ganeev_pra85}%
  \BibitemOpen
  \bibfield  {author} {\bibinfo {author} {\bibfnamefont {R.~A.}\ \bibnamefont
  {Ganeev}}, \bibinfo {author} {\bibfnamefont {V.~V.}\ \bibnamefont
  {Strelkov}}, \bibinfo {author} {\bibfnamefont {C.}~\bibnamefont {Hutchison}},
  \bibinfo {author} {\bibfnamefont {A.}~\bibnamefont {Za\"{\i}r}}, \bibinfo
  {author} {\bibfnamefont {D.}~\bibnamefont {Kilbane}}, \bibinfo {author}
  {\bibfnamefont {M.~A.}\ \bibnamefont {Khokhlova}},\ and\ \bibinfo {author}
  {\bibfnamefont {J.~P.}\ \bibnamefont {Marangos}},\ }\bibfield  {title}
  {\bibinfo {title} {Experimental and theoretical studies of two-color-pump
  resonance-induced enhancement of odd and even harmonics from a tin plasma},\
  }\href {https://doi.org/10.1103/PhysRevA.85.023832} {\bibfield  {journal}
  {\bibinfo  {journal} {Phys. Rev. A}\ }\textbf {\bibinfo {volume} {85}},\
  \bibinfo {pages} {023832} (\bibinfo {year} {2012})}\BibitemShut {NoStop}%
\bibitem [{\citenamefont {Corkum}(1993)}]{3-step_C}%
  \BibitemOpen
  \bibfield  {author} {\bibinfo {author} {\bibfnamefont {P.~B.}\ \bibnamefont
  {Corkum}},\ }\bibfield  {title} {\bibinfo {title} {Plasma perspective on
  strong field multiphoton ionization},\ }\href
  {https://doi.org/10.1103/PhysRevLett.71.1994} {\bibfield  {journal} {\bibinfo
   {journal} {Phys. Rev. Lett.}\ }\textbf {\bibinfo {volume} {71}},\ \bibinfo
  {pages} {1994} (\bibinfo {year} {1993})}\BibitemShut {NoStop}%
\bibitem [{\citenamefont {Schafer}\ \emph {et~al.}(1993)\citenamefont
  {Schafer}, \citenamefont {Yang}, \citenamefont {DiMauro},\ and\ \citenamefont
  {Kulander}}]{3-step_S}%
  \BibitemOpen
  \bibfield  {author} {\bibinfo {author} {\bibfnamefont {K.~J.}\ \bibnamefont
  {Schafer}}, \bibinfo {author} {\bibfnamefont {B.}~\bibnamefont {Yang}},
  \bibinfo {author} {\bibfnamefont {L.~F.}\ \bibnamefont {DiMauro}},\ and\
  \bibinfo {author} {\bibfnamefont {K.~C.}\ \bibnamefont {Kulander}},\
  }\bibfield  {title} {\bibinfo {title} {Above threshold ionization beyond the
  high harmonic cutoff},\ }\href {https://doi.org/10.1103/PhysRevLett.70.1599}
  {\bibfield  {journal} {\bibinfo  {journal} {Phys. Rev. Lett.}\ }\textbf
  {\bibinfo {volume} {70}},\ \bibinfo {pages} {1599} (\bibinfo {year}
  {1993})}\BibitemShut {NoStop}%
\bibitem [{\citenamefont {Kuchiev}(1987)}]{3-step_K}%
  \BibitemOpen
  \bibfield  {author} {\bibinfo {author} {\bibfnamefont {M.~Y.}\ \bibnamefont
  {Kuchiev}},\ }\bibfield  {title} {\bibinfo {title} {Atomic antenna},\
  }\href@noop {} {\bibfield  {journal} {\bibinfo  {journal} {Pis'ma Zh. Eksp.
  Teor. Fiz.}\ }\textbf {\bibinfo {volume} {45}},\ \bibinfo {pages} {319}
  (\bibinfo {year} {1987})},\ \bibinfo {note} {[JETP Lett. {\bf 45}, 404
  (1987)]}\BibitemShut {NoStop}%
\bibitem [{\citenamefont {Strelkov}\ \emph {et~al.}(2012)\citenamefont
  {Strelkov}, \citenamefont {Khokhlova}, \citenamefont {Gonoskov},
  \citenamefont {Gonoskov},\ and\ \citenamefont {Ryabikin}}]{Strelkov_PRA_EPL}%
  \BibitemOpen
  \bibfield  {author} {\bibinfo {author} {\bibfnamefont {V.~V.}\ \bibnamefont
  {Strelkov}}, \bibinfo {author} {\bibfnamefont {M.~A.}\ \bibnamefont
  {Khokhlova}}, \bibinfo {author} {\bibfnamefont {A.~A.}\ \bibnamefont
  {Gonoskov}}, \bibinfo {author} {\bibfnamefont {I.~A.}\ \bibnamefont
  {Gonoskov}},\ and\ \bibinfo {author} {\bibfnamefont {M.~Y.}\ \bibnamefont
  {Ryabikin}},\ }\bibfield  {title} {\bibinfo {title} {High-order harmonic
  generation by atoms in an elliptically polarized laser field: Harmonic
  polarization properties and laser threshold ellipticity},\ }\href
  {https://doi.org/10.1103/PhysRevA.86.013404} {\bibfield  {journal} {\bibinfo
  {journal} {Phys. Rev. A}\ }\textbf {\bibinfo {volume} {86}},\ \bibinfo
  {pages} {013404} (\bibinfo {year} {2012})}\BibitemShut {NoStop}%
\bibitem [{\citenamefont {Ammosov}\ \emph {et~al.}(1986)\citenamefont
  {Ammosov}, \citenamefont {Delone},\ and\ \citenamefont {Krainov}}]{ADK}%
  \BibitemOpen
  \bibfield  {author} {\bibinfo {author} {\bibfnamefont {M.~V.}\ \bibnamefont
  {Ammosov}}, \bibinfo {author} {\bibfnamefont {N.~B.}\ \bibnamefont
  {Delone}},\ and\ \bibinfo {author} {\bibfnamefont {V.~P.}\ \bibnamefont
  {Krainov}},\ }\bibfield  {title} {\bibinfo {title} {Tunnel ionization of
  complex atoms and of atomic ions in an alternating electromagnetic field},\
  }\href@noop {} {\bibfield  {journal} {\bibinfo  {journal} {Zh. Eksp. Teor.
  Fiz.}\ }\textbf {\bibinfo {volume} {91}},\ \bibinfo {pages} {2008} (\bibinfo
  {year} {1986})},\ \bibinfo {note} {[Sov. Phys. JETP {\bf 64}, 1191
  (1986)]}\BibitemShut {NoStop}%
\bibitem [{\citenamefont {Perelomov}\ \emph {et~al.}(1966)\citenamefont
  {Perelomov}, \citenamefont {Popov},\ and\ \citenamefont {Terent'ev}}]{PPT}%
  \BibitemOpen
  \bibfield  {author} {\bibinfo {author} {\bibfnamefont {A.~M.}\ \bibnamefont
  {Perelomov}}, \bibinfo {author} {\bibfnamefont {V.~S.}\ \bibnamefont
  {Popov}},\ and\ \bibinfo {author} {\bibfnamefont {M.~V.}\ \bibnamefont
  {Terent'ev}},\ }\bibfield  {title} {\bibinfo {title} {Ionization of atoms in
  an alternating electric field},\ }\href@noop {} {\bibfield  {journal}
  {\bibinfo  {journal} {Zh. Eksp. Teor. Fiz.}\ }\textbf {\bibinfo {volume}
  {50}},\ \bibinfo {pages} {1393} (\bibinfo {year} {1966})},\ \bibinfo {note}
  {[Sov. Phys. JETP {\bf 23}, 924 (1966)]}\BibitemShut {NoStop}%
\bibitem [{\citenamefont {Clementi}\ \emph {et~al.}(1967)\citenamefont
  {Clementi}, \citenamefont {Raimondi},\ and\ \citenamefont
  {Reinhardt}}]{Clementi1967Aug}%
  \BibitemOpen
  \bibfield  {author} {\bibinfo {author} {\bibfnamefont {E.}~\bibnamefont
  {Clementi}}, \bibinfo {author} {\bibfnamefont {D.~L.}\ \bibnamefont
  {Raimondi}},\ and\ \bibinfo {author} {\bibfnamefont {W.~P.}\ \bibnamefont
  {Reinhardt}},\ }\bibfield  {title} {\bibinfo {title} {{Atomic screening
  constants from SCF functions. II. Atoms with 37 to 86 electrons}},\ }\href
  {https://doi.org/10.1063/1.1712084} {\bibfield  {journal} {\bibinfo
  {journal} {J. Chem. Phys.}\ }\textbf {\bibinfo {volume} {47}},\ \bibinfo
  {pages} {1300} (\bibinfo {year} {1967})}\BibitemShut {NoStop}%
\bibitem [{\citenamefont {Strelkov}\ \emph {et~al.}(2011)\citenamefont
  {Strelkov}, \citenamefont {Gonoskov}, \citenamefont {Gonoskov},\ and\
  \citenamefont {Ryabikin}}]{Ell_Origin}%
  \BibitemOpen
  \bibfield  {author} {\bibinfo {author} {\bibfnamefont {V.~V.}\ \bibnamefont
  {Strelkov}}, \bibinfo {author} {\bibfnamefont {A.~A.}\ \bibnamefont
  {Gonoskov}}, \bibinfo {author} {\bibfnamefont {I.~A.}\ \bibnamefont
  {Gonoskov}},\ and\ \bibinfo {author} {\bibfnamefont {M.~Y.}\ \bibnamefont
  {Ryabikin}},\ }\bibfield  {title} {\bibinfo {title} {Origin for ellipticity
  of high-order harmonics generated in atomic gases and the sublaser-cycle
  evolution of harmonic polarization},\ }\href
  {https://doi.org/10.1103/PhysRevLett.107.043902} {\bibfield  {journal}
  {\bibinfo  {journal} {Phys. Rev. Lett.}\ }\textbf {\bibinfo {volume} {107}},\
  \bibinfo {pages} {043902} (\bibinfo {year} {2011})}\BibitemShut {NoStop}%
\bibitem [{\citenamefont {Fareed}\ \emph {et~al.}(2017)\citenamefont {Fareed},
  \citenamefont {Strelkov}, \citenamefont
  {Thir{\ifmmode\acute{e}\else\'{e}\fi}}, \citenamefont {Mondal}, \citenamefont
  {Schmidt}, \citenamefont
  {L{\ifmmode\acute{e}\else\'{e}\fi}gar{\ifmmode\acute{e}\else\'{e}\fi}},\ and\
  \citenamefont {Ozaki}}]{Fareed2017}%
  \BibitemOpen
  \bibfield  {author} {\bibinfo {author} {\bibfnamefont {M.~A.}\ \bibnamefont
  {Fareed}}, \bibinfo {author} {\bibfnamefont {V.~V.}\ \bibnamefont
  {Strelkov}}, \bibinfo {author} {\bibfnamefont {N.}~\bibnamefont
  {Thir{\ifmmode\acute{e}\else\'{e}\fi}}}, \bibinfo {author} {\bibfnamefont
  {S.}~\bibnamefont {Mondal}}, \bibinfo {author} {\bibfnamefont {B.~E.}\
  \bibnamefont {Schmidt}}, \bibinfo {author} {\bibfnamefont {F.}~\bibnamefont
  {L{\ifmmode\acute{e}\else\'{e}\fi}gar{\ifmmode\acute{e}\else\'{e}\fi}}},\
  and\ \bibinfo {author} {\bibfnamefont {T.}~\bibnamefont {Ozaki}},\ }\bibfield
   {title} {\bibinfo {title} {{High-order harmonic generation from the dressed
  autoionizing states}},\ }\href {https://doi.org/10.1038/ncomms16061}
  {\bibfield  {journal} {\bibinfo  {journal} {Nat. Commun.}\ }\textbf {\bibinfo
  {volume} {8}},\ \bibinfo {pages} {1} (\bibinfo {year} {2017})}\BibitemShut
  {NoStop}%
\bibitem [{\citenamefont {Ganeev}\ \emph {et~al.}(2014)\citenamefont {Ganeev},
  \citenamefont {Abdelrahman}, \citenamefont {Frank}, \citenamefont {Witting},
  \citenamefont {Okell}, \citenamefont {Fabris}, \citenamefont {Hutchison},
  \citenamefont {Marangos},\ and\ \citenamefont {Tisch}}]{Ganeev2014}%
  \BibitemOpen
  \bibfield  {author} {\bibinfo {author} {\bibfnamefont {R.~A.}\ \bibnamefont
  {Ganeev}}, \bibinfo {author} {\bibfnamefont {Z.}~\bibnamefont {Abdelrahman}},
  \bibinfo {author} {\bibfnamefont {F.}~\bibnamefont {Frank}}, \bibinfo
  {author} {\bibfnamefont {T.}~\bibnamefont {Witting}}, \bibinfo {author}
  {\bibfnamefont {W.~A.}\ \bibnamefont {Okell}}, \bibinfo {author}
  {\bibfnamefont {D.}~\bibnamefont {Fabris}}, \bibinfo {author} {\bibfnamefont
  {C.}~\bibnamefont {Hutchison}}, \bibinfo {author} {\bibfnamefont {J.~P.}\
  \bibnamefont {Marangos}},\ and\ \bibinfo {author} {\bibfnamefont {J.~W.~G.}\
  \bibnamefont {Tisch}},\ }\bibfield  {title} {\bibinfo {title} {{Spatial
  coherence measurements of non-resonant and resonant high harmonics generated
  in laser ablation plumes}},\ }\href {https://doi.org/10.1063/1.4861161}
  {\bibfield  {journal} {\bibinfo  {journal} {Appl. Phys. Lett.}\ }\textbf
  {\bibinfo {volume} {104}},\ \bibinfo {pages} {021122} (\bibinfo {year}
  {2014})}\BibitemShut {NoStop}%
\bibitem [{\citenamefont {Strelkov}(2016)}]{Strelkov2016}%
  \BibitemOpen
  \bibfield  {author} {\bibinfo {author} {\bibfnamefont {V.~V.}\ \bibnamefont
  {Strelkov}},\ }\bibfield  {title} {\bibinfo {title} {{Attosecond-pulse
  production using resonantly enhanced high-order harmonics}},\ }\href
  {https://doi.org/10.1103/PhysRevA.94.063420} {\bibfield  {journal} {\bibinfo
  {journal} {Phys. Rev. A}\ }\textbf {\bibinfo {volume} {94}},\ \bibinfo
  {pages} {063420} (\bibinfo {year} {2016})}\BibitemShut {NoStop}%
\end{thebibliography}%

\end{document}